\documentclass[twocolumn]{openjournal}

\usepackage[T1]{fontenc}
\usepackage{ae,aecompl}

\usepackage{graphicx}	
\usepackage{amsmath}	
\usepackage{amssymb}	
\usepackage{natbib}

\usepackage{hyperref}
\hypersetup{
	colorlinks=true,
	urlcolor=blue,    
	linkcolor=black,  
	citecolor=blue,   
}

\defcitealias{Pittordis_2018}{PS18}
\defcitealias{Pittordis_2019}{PS19}
\defcitealias{Badry_2021}{ERH}
\defcitealias{Banik_2018_Centauri}{BZ18}

\newcommand{\draftdate}{ Accepted by Open Journal of Astrophysics, 31 Jan 2023; original submission 13 July 2022 } 
\newcommand{\kms} {{\rm \, km \, s^{-1} }} 

\newcommand{\au} {\, {\rm AU}}   
\newcommand{\kau} {\, {\rm kAU}} 
\newcommand{\msun} {\,M_\odot} 
 
\newcommand{\pc} {\, {\rm pc}} 
\newcommand{\kpc} {\, {\rm kpc}}

\newcommand{\vp} {v_{\rm p}} 
\newcommand{\vc}{v_{\rm C}} 
\newcommand{\vtilde}{\tilde{v}} 
\newcommand{\vinf}{v_\infty} 
\newcommand{\rp} {r_{\rm p}} 
\newcommand{\mas} {\, {\rm mas}} 
\newcommand{\vratio}{\vp / \vc(\rp)} 
 \newcommand{\newa}{  }  

\begin{document}
	
	\title{Wide Binaries from GAIA EDR3: preference for GR over MOND ?\\}
	\date{\draftdate}
	
	\author{C. Pittordis} 
	\email[E-mail: ]{cp.pittordis@gmail.com}
	\author{W. Sutherland} 
	\email[E-mail: ]{w.j.sutherland@qmul.ac.uk}
	\affiliation{The School of Physical and Chemical Sciences,  
		Queen Mary University of London, Mile End Road, London E1 4NS, UK.}

	\begin{abstract}
		Several recent studies have shown that velocity differences of very wide binary stars, measured to high precision with GAIA, can potentially provide an interesting test for modified-gravity theories which attempt to emulate dark matter. These systems should be entirely Newtonian according to standard dark-matter theories, while the predictions for MOND-like theories are distinctly different, if the various observational issues can be overcome. Here we provide an updated version of our 2019 study using the recent GAIA EDR3 data release: we select a large sample of 73\,159 candidate wide binary stars with distance $\leq$300 parsec and magnitudes G$<$17 from GAIA EDR3, and estimate component masses using a main-sequence mass-luminosity relation. We then examine the frequency distribution of pairwise relative projected velocity (relative to circular-orbit value) as a function of projected separation, compared to simulations; as before, these distributions show a clear peak at a value close to Newtonian expectations, along with a long `tail' which extends to much larger velocity ratios and may well be caused by hierarchical triple systems with an unresolved or unseen third star.  
		We then fit these observed distributions with a simulated mixture of binary, triple and flyby populations, 
		for GR or MOND orbits, and find that standard gravity is somewhat preferred over one specific implementation of MOND; 
		though we have not yet explored the full parameter space of triple population models and MOND versions.  Improved data from future GAIA releases, and followup of a subset of systems to better characterise the triple population, should allow wide binaries to become a decisive test of GR vs MOND in the future. 
	\end{abstract}
	\maketitle
	
	\section{Introduction}
	
	Einstein's theory of General Relativity (GR) is perceived to be the best description of gravity on all scales. GR has been very well tested in the Solar System, but requires exotic dark matter (DM) to explain larger structures such as galaxies, clusters, and the cosmic microwave background (CMB) within the GR-based $\Lambda$CDM cosmological model \citep{Planck_2015}. At present time, there is no decisive direct detection of DM (e.g. \citet{LUX_2016}); though there are many lines of indirect evidence, primarily via flat rotation curves (RCs) of galaxies, gravitational lensing, galaxy clustering and the CMB. Moreover, the $\Lambda$CDM picture leads to several small-scale discrepancies between numerical simulations based on DM and observations on sub-galactic scales \citep{Kroupa_2015, Bullock_2017}, perhaps most seriously with the Local Group satellite planes \citep{Pawlowski_2014}.
	
	This leaves room for modified gravity (MG) theories such as the notable MOdified Newtonian Dynamics (MOND) \citep{Milgrom_1983} as an alternative to GR.  MOND \citep[reviewed in][]{Famaey_McGaugh_2012} attempts to explain 
	the observations usually attributed to dark matter by a suitable modification of standard GR.  
	The original MOND formulation was non-relativistic and really a
	fitting function rather than a realistic theory; 
	it has later been incorporated into relativistic theories following from
	the well-known Tensor-Vector-Scalar (TeVeS) theory proposed by
	\citet{Bekenstein_2004}. There are various versions of MG, but figure 10 from \citet{Famaey_McGaugh_2012} indicates that the most promising should involve a departure from Newtonian gravity at accelerations below some threshold $a_{_0}$, as occurs in MOND.
	The current situation of $\Lambda$CDM vs MOND remains contentious, with many successes for $\Lambda$CDM on large scales but some anomalies on galactic and sub-galactic scales; while MOND appears mostly successful  on galactic scales but has 
	no clear-cut predictions for cosmology.  
	
	As noted in \citet[][hereafter \citetalias{Pittordis_2019}]{Pittordis_2019}, a clear convincing direct detection of dark matter
	in underground experiments would be the most decisive scenario, but the converse is not 
	true: null results from dark matter detectors can never
	rule out the paradigm, because the DM interaction cross-section might simply be too small for any practical experiment (or, the cross-section could be weak-like but the DM particle masses could be very large $\sim 10^9\,$GeV, implying a local number density far below the value for conventional TeV-scale WIMPs). Therefore, in the absence of a DM direct detection,
	new tests which can discriminate between DM and modified gravity from direct tests of gravity at the relevant very low
	accelerations are highly desirable.
	
	A promising test of MG is via wide binary (WB) stellar systems in the Solar neighbourhood \citep[within $\approx 250$~pc,][]{Hernandez_2012}. WBs are isolated stellar pairs with separations $\ga 7 \kau$ in order to have orbital accelerations $\la a_{_0} = 1.2 \times 10^{-10}$~m/s\textsuperscript{2}. 
	The small size of WBs means no significant amount of DM should be distributed within them, so Keplerian dynamics should apply in GR+DM models. 
	In a MOND context, local WBs must feel the Galactic external field which is slightly larger than $a_0$, and the external
	field effect results in 
	WB orbital velocities exceeding the Newtonian prediction by $\sim 20\%$ \citep{Banik_2018_Centauri} for an interpolating function consistent with galaxy RCs \citep{McGaugh_Lelli_2016}. This excess $\approx 20\%$ remains approximately constant at larger WB separations, although WB tests are more prone to observational 
	caveats at larger separations. Studies of WBs in general have been explored by e.g. 
	\citet{Weinberg_1987},
	\citet{Close_1990},
	\citet{Yoo_2003},
	\citet{Lepine_2007},
	\citet{Kouwenhoven_2010},
	\citet{Jiang_2010},
	\citet{Dhital_2013}, 
	\citet{Coronado_2018}, and others. 
	
	Previous work concerning
	tests of MOND-like gravity has been done by
	\citet{Hernandez_2012},
	\citet{Hernandez_2012_2}
	\citet{Hernandez_2014},
	\citet{Matvienko_2015}, 
	\citet{Scarpa_2017} 
	and \citet{Hernandez_2019}; these typically give hints of deviations in
	the direction expected from MOND-like gravity, though due
	to the limited precision of the pre-GAIA data used, these hints
	were not decisive. If modelled appropriately, WBs can provide a very `clean' model-independent test of gravity in the extremely weak-field regime relevant to galactic outskirts and large scale structure. 
	Note that the orbital periods of WBs are $\sim$ Myr, so accelerations are undetectable and full orbit modelling is not possible, thus each single binary is strongly under-constrained; however the statistical distribution of relative velocities is readily predicted from simulations assuming random phases and inclinations, so a large sample of thousands of WBs  can provide a sensitive test.  
	
	In our first paper in this series \citep[][hereafter \citetalias{Pittordis_2018}]{Pittordis_2018}, we 
	used simulations to explore the prospects for
	this wide-binary test, in anticipation of the much improved data 
	from the GAIA spacecraft \citep{Gaia_2016}; \citetalias{Pittordis_2018} used
	simulated wide-binary orbits for a variety of acceleration
	laws, including Newtonian and various MOND models both
	with and without an external field effect (hereafter ExFE);
	the general conclusion was that GAIA data provides promising prospects for such a test, since MOND-like models without an ExFE should produce large and obvious deviations
	towards larger velocity differences.
	In MOND-like models with the ExFE included, as theoretically preferred, the local Galactic acceleration field substantially suppresses MOND-like effects, but does not eliminate
	them. These models with ExFE give predicted relative velocities much
	closer to Newtonian, but do still show subtle deviations,
	most notably a significantly larger fraction of binaries with
	pairwise velocities in the range $(1.1 - 1.5) \times \vc(\rp)$, where
	$\vc(\rp)$ is the Newtonian circular velocity at projected separation $\rp$. 
	Here we recall two of the main conclusions from \citetalias{Pittordis_2018}:
	MOND-like theories can allow bound binaries with relative
	velocities above the Newtonian ceiling, $v_{3D}/\vc(\rp) >\sqrt{2}$,
	(where $v_{3D}$ is the 3-D pairwise relative velocity); but with
	the ExFE included the fraction of such systems is predicted
	to be very small, typically 1 percent or less; so simply counting such systems is unlikely to be a practical test due to possible contamination, observational errors, and small-number
	statistics. However, the upper percentiles of this velocity ratio, or similarly the fraction of binaries with velocity ratio
	between $\sim 1.1 - 1.5$ are more promising statistics.  {\newa In particular, the upper tail (top $\sim 10$ percent)
		of the binary velocity distribution is only weakly sensitive to the rather uncertain eccentricity distribution, but is strongly
		sensitive to modified gravity. }  
	
	The second main conclusion from \citetalias{Pittordis_2018} was that, in
	MOND theories including the ExFE, there is an optimal window
	of binary projected separations, $5 \la \rp \la 20 \kau$, for practical application of the test. 
	Even wider separations are not favoured
	in practice because the inclusion of the ExFE causes the
	MOND-like effects to almost saturate at a fixed percentage at $\rp \ga 10 \kau$, while
	several observational issues become proportionally worse at even wider separations $\ga 20 \kau$.

	After the release of GAIA Data Release 2, 
	we explored an observational application of the test with that dataset  \citep[][hereafter PS19]{Pittordis_2019} 
	{\newa (see also \citet{Hernandez_2019} and \citet{Hernandez_2022} for related studies with GAIA data) }.  
	Since the 3D separation $r$ is not a practical observable (since the line-of-sight separation is typically well below
	the precision of distance measurements, even for GAIA) we have to replace
	it with projected separation $r_p$ as a proxy, which shifts the
	ratios to lower values depending on viewing angles; but this
	effect can be readily included in simulations.   The main conclusions of PS19 were that 
	there is a sufficient population of bound binaries to carry out the test in future, but also that
	there exists a ``tail" of pairs with large velocity differences, too numerous to explain by random alignments. 
	This may be accounted for by common-origin unbound pairs (as modelled in PS19), or perhaps a more likely
	explanation by \cite{Clarke_2020} is hierarchical triples with an unresolved or faint third star in a closer orbit, 
	which boosts the observed velocity difference between the two visible stars.  
	
	In this paper we provide an updated version of PS19, using the recent GAIA Early Data Release 3 \citep{GaiaEDR3_2020a, GaiaEDR3_2020b} to provide a $\sim 3 \times $ larger sample of candidate WBs, with more precise proper motions.  
	We also update our modelling here to fit the observed
	distributions with a mixture of binary, triple and flyby systems, using both Newtonian and MOND gravity laws and 
	various eccentricity distributions.

	The plan of this paper is as follows: in Section 2 we
	describe the methods for selecting candidate wide-binaries from the
	GAIA Early Data Release 3 
	(hereafter, EDR3) data and cleaning the sample; we then compare to the earlier sample from \citet{Pittordis_2019}, 
	and the EDR3 sample from  \citet{Badry_2021}.  In Section 3 we discuss various simulations of the velocity-ratio distributions
	for Newtonian and the realistic MOND-like binary orbits (with ExFE) from \citep[][hereafter \citetalias{Banik_2018_Centauri}]{Banik_2018_Centauri}; and we
	also simulate triple star systems, which appear to provide a reasonable fit to the high-velocity tail of the distribution, $\vratio \geq \sqrt2$, and unbound flyby systems.   In Section 4 we fit the observed distributions with a mixture of simulated binary,  triple and flyby model populations, for both GR and MOND cases; and
	we summarise our conclusions in Section 5.
	
	\section{GAIA EDR3 and sample selection} 
	\label{sec:edr3} 
	
	\subsection{Preliminary selection} 
	\label{sec:preliminary selection}
	
	Our starting point is the public GAIA Early Data Release 3 dataset (EDR3),
	\citep{GaiaEDR3_2020a, GaiaEDR3_2020b} released on 2020 December 3. 
	We initially select all stars with measured parallax $\omega > \frac{10}{3} \mas$
	(i.e. estimated distance $< 300 \pc$) with a GAIA broadband magnitude $G < 17$, and cutting out the Galactic plane with absolute latitude $\vert b \vert \le 15 \deg$ yielding a preliminary EDR3 sample of 2,101,920 stars (hereafter PEDR3). 
	(Data quality cuts are applied at a later stage, in order that these
	may be adjusted post-selection).  These criteria are expanded from $200 \pc$ and $G < 16$ used by 
	\citetalias{Pittordis_2019} from Gaia Data Release 2 (hereafter, DR2), since the improved data quality in EDR3 allows
	an expanded volume while giving slightly smaller transverse velocity errors. 
	The parallax and magnitude cuts above are chosen to provide a 
	large enough volume to contain a 
	usefully large statistical sample of wide binaries; while the moderate
	distance limit and relatively bright magnitude limit ensures that 
	GAIA provides high precision on distances and transverse velocities.
	
	We then used the same search method as described in section 2 from \citetalias{Pittordis_2019} to search this nearby-star sample for pairs of stars
	with projected separation $\le 50 \kau$ (calculated 
	at the mean distance of each candidate pair), 
	difference in parallax distances of both stars $\vert d_1 - d_2 \vert \leq \, 8.0 \pc$, and distances also 
	consistent with each other within $4\times$
	the combined uncertainty e.g., $ \vert d_1 - d_2 \vert \leq 4\sigma_{d}$,  and projected velocity difference $\Delta v_p \le 3 \kms$ 
	as inferred from the difference in proper motions; here, the projected velocity difference is computed  
	assuming {\em both} stars in each candidate pair are
	actually at the mean of the two estimated distances. 
	
	We note here that this common-distance assumption is important:
	if the relative velocities are calculated using 
	individual parallax distances, then an example random 1 percent
	difference in parallax for a system
	with transverse velocity $40 \kms$ scales to a $0.4 \kms$ transverse 
	velocity difference, which is similar to or larger than the orbital
	velocities of interest below. 
	However, since we are almost entirely 
	interested in the velocity {\em difference} 
	within a binary, the common-distance assumption leads to an error
	in estimated relative velocity proportional to 
	the unknown {\em true} fractional distance difference, $(d_1 - d_2)/ d$ 
	(see also \citet{Shaya_2011}, and Section 2.4 of \citetalias{Pittordis_2018} 
	for related perspective effects).   
	For true binaries with random orientation we 
	expect $ \vert d_1 - d_2 \vert \le \rp$ for 71 percent of systems, 
	and $\le 2 \rp$ for 90 percent.  
	Then for a typical binary (see below) 
	with $\rp \sim 10 \kau$ and $d \sim 180 \pc$
	we have $\rp/d \sim 2.7 \times 10^{-4}$; this is 
	much smaller than the fractional uncertainty of the parallaxes,
	so choosing the common-distance assumption yields a much more 
	precise estimate of the relative velocity for {\em genuine} binaries. 
	
\begin{figure*}
		\begin{center} 
			\includegraphics[width=\linewidth]{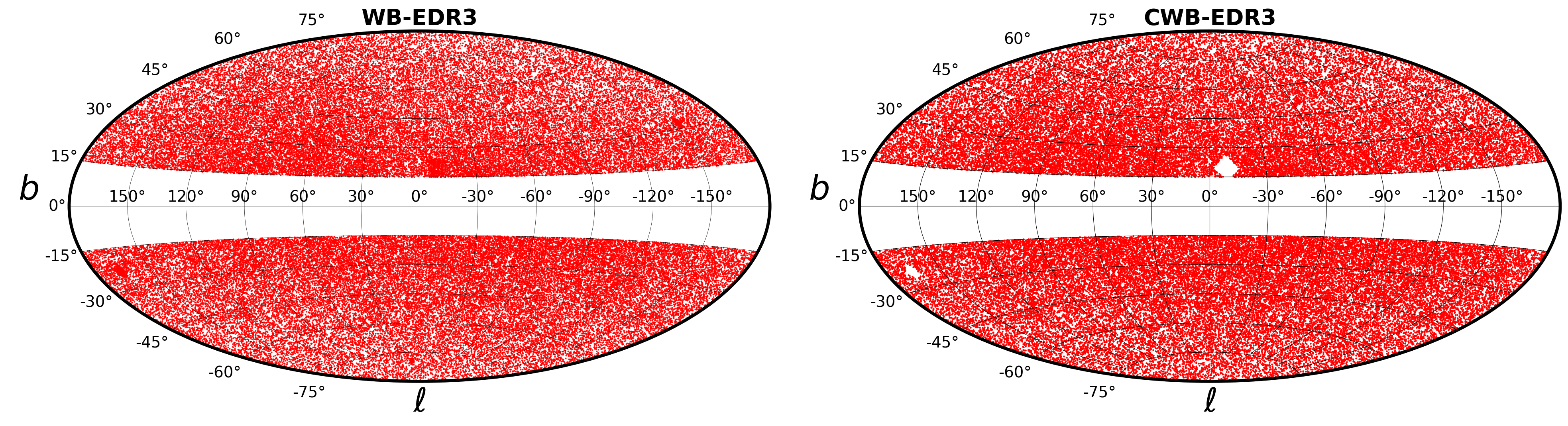} 
			\caption{ 
				The sky distribution in Galactic coordinates for candidate
				binaries in EDR3: the left panel shows the initial set of candidate binaries WB-EDR3, the right panel is 
				the cleaned sample CWB-EDR3 after applying all cuts in Section~\ref{sec:edr3}.  Note that the holes from 
				limited sky coverage in PS19 no longer appear with the improved coverage of EDR3.  } 
			\label{fig:aitoff} 
		\end{center} 
	\end{figure*}

	This search results in a first-cut
	sample of 92,677 candidate EDR3 wide binaries (hereafter WB-EDR3); this sample is then pruned with
	additional cuts as described in the following subsections.
	The sky distribution of these candidates is shown in Figure~\ref{fig:aitoff} (left panel). 
	
	\subsection{Triple and higher systems} 
	\label{sec:triples} 
	To reject the majority of ``moving groups'' or similar,
	we searched our WB-EDR3 sample for any star in common between two 
	or more candidate binaries; if so, both or all those binaries 
	were rejected,  reducing the WB-EDR3 sample to 83,884 candidate binaries with no 
	star common to more than one candidate binary. 
	
	\subsection{Sky cuts} 
	\label{sec:skycut}
	Inspection of the WB-EDR3 sample showed a roughly uniform
	distribution across the sky,  
	with some enhancement near around some well-known open clusters, 
	i.e. the Pleiades, Praesepe, Upper Sco, and the Blanco 1 open cluster.  
	Similar to \citetalias{Pittordis_2019}, we applied sky cuts to eliminate regions around the above clusters; 
	the details of the exclusion regions are listed in Table~\ref{tab:openclus}.  
	These sky cuts reduce the sample to 83,100 candidate binaries. 
	
	\begin{table} 
		\caption{ List of coordinates for excluded regions around open clusters.} 
		\centering
		\begin{tabular}{lcc} 
			\hline
			Cluster  & RA limits (deg) & Dec limits (deg) \\ 
			\hline 
			Blanco 1 & (0, 2) & (-31,  -29) \\ 
			Pleiades & (54,  59) & (+22,  +27) \\
			Praesepe & (129, 131) & (+18, +22) \\ 
			Upper Sco & (238, 248) & (-29, -19) \\
			\hline 
		\end{tabular} 
		\label{tab:openclus} 
	\end{table}

	\subsection{Removing faint companions}
	\label{sec:faintstars}
	Similar to the methods of \citetalias{Pittordis_2019}, we also searched for additional co-moving companion stars to a fainter limit: we selected a ``faint star" sample of GAIA stars with $G < 20$ and a measured parallax $\omega > \frac{10}{3} \mas$
	(i.e. estimated distance $< 300 \pc$); for each star in each candidate binary, we then searched for faint-star companions with the following criteria: 
	\begin{enumerate} 
		\item Parallax consistent with the main star at $4\sigma$. 
		\item Angular separation less than $2/3$ of the main-binary separation 
		(since hierarchical triples are expected to be unstable for inner-orbit
		separation above $\sim 0.4 \times$ the outer separation); 
		and angular separation above $0.5$ arcsec to avoid barely-resolved
		companions.  
		\item Measured projected velocity difference from the main star 
		$\le 5 \kms$. 
	\end{enumerate} 
	If any such ``third star'' was found, (in 698 cases), we rejected the 
	candidate binary since a hierarchical triple will generally
	boost the projected velocity difference of the wide pair; 
	this left a de-tripled sample of 82,402 candidate binaries from the WB-EDR3 sample.  
	
	Clearly, the third-star search above will not reject third stars
	which are either very faint or unresolved from one of our 
	binary members;  this will need to be considered for possible 
	follow-up observations later. We consider the effects of contamination from undetected triple systems later in section \ref{sec:triple-sim}, showing this may be the dominant source of contamination.
	
	\subsection{Data quality cuts on wide binary candidates}
	\label{sec: Data quality cuts}
	We applied data-quality cuts to the WB-EDR3 sample based on the GAIA 
	parameters, as \citet{Arenou_2018}  Equation 1 as follows: 
	\begin{eqnarray} 
		\chi^2 & \equiv  & {\tt astrometric\_chi2\_al}  \nonumber \\
		\nu & \equiv & {\tt astrometric\_n\_good\_obs\_al } - 5 \nonumber \\
		u & \equiv & \sqrt{\chi^2 / \nu} \nonumber \\  
		u & \le & 1.2 \times {\rm max}(1, \exp\left[ -0.2(G-19.5) \right] ) 
		\label{eq:arenou} 
	\end{eqnarray} 
	
	We rejected candidate pairs from the WB-EDR3 sample, where at least one star did not satisfy Eq.~\ref{eq:arenou}; 
	this rejected 9,243 candidate pairs, leaving a cleaned sample of 73,159 candidate wide binaries,
	hereafter labelled CWB-EDR3: this is the main sample used for analysis below. 

	\subsection{Results and scaled velocities} 
	
\begin{figure*} 
		\begin{center} 
			\includegraphics[width=16cm]{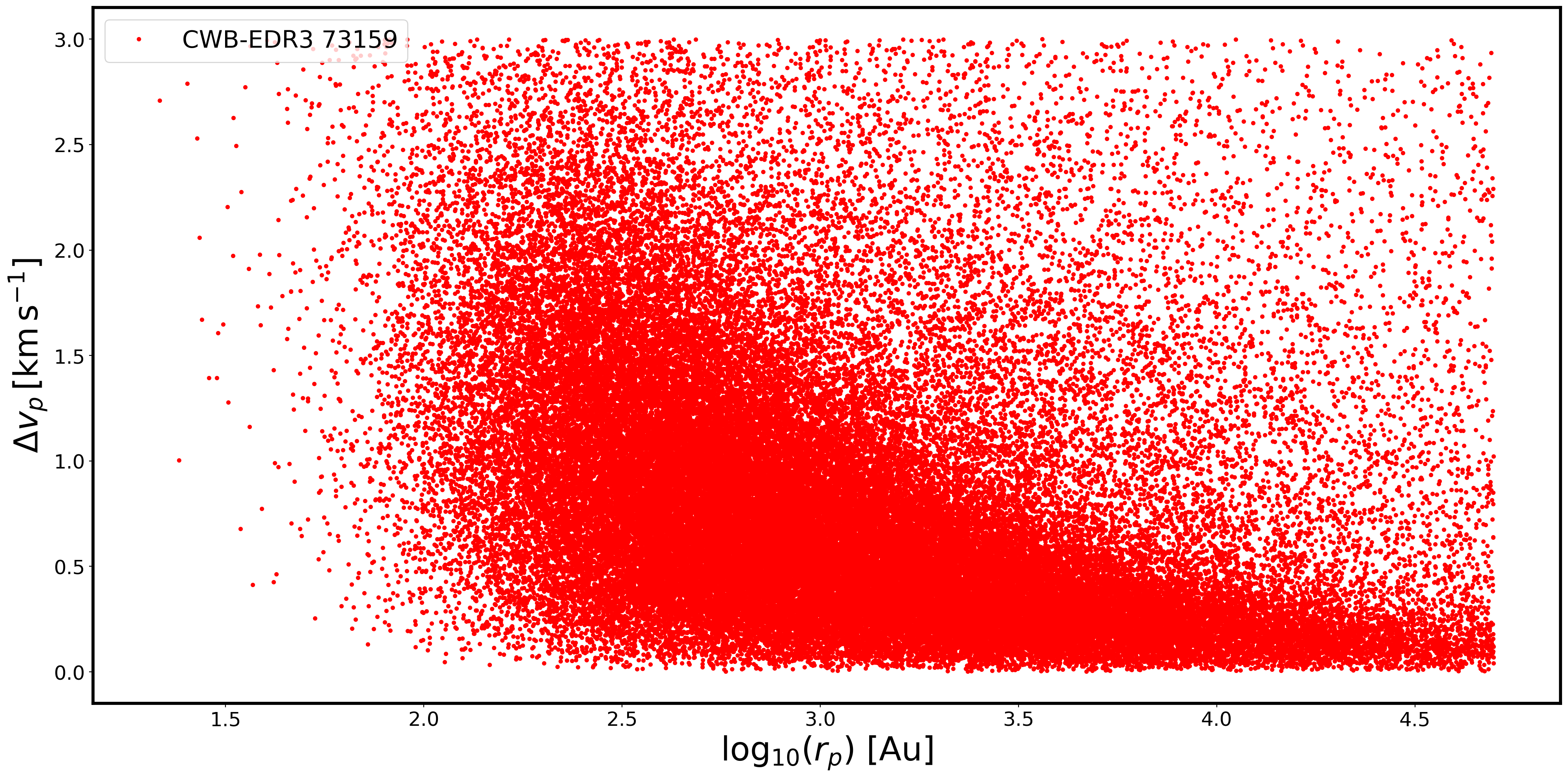} 
			\caption{Scatter plot of projected relative velocity $\vp$ (y-axis)
				vs projected separation (log scale, x-axis) for the CWB-EDR3 binary candidates.  
				The main selection cuts are visible at top and right. } 
			\label{fig:rpvp} 
			
			\includegraphics[width=16cm]{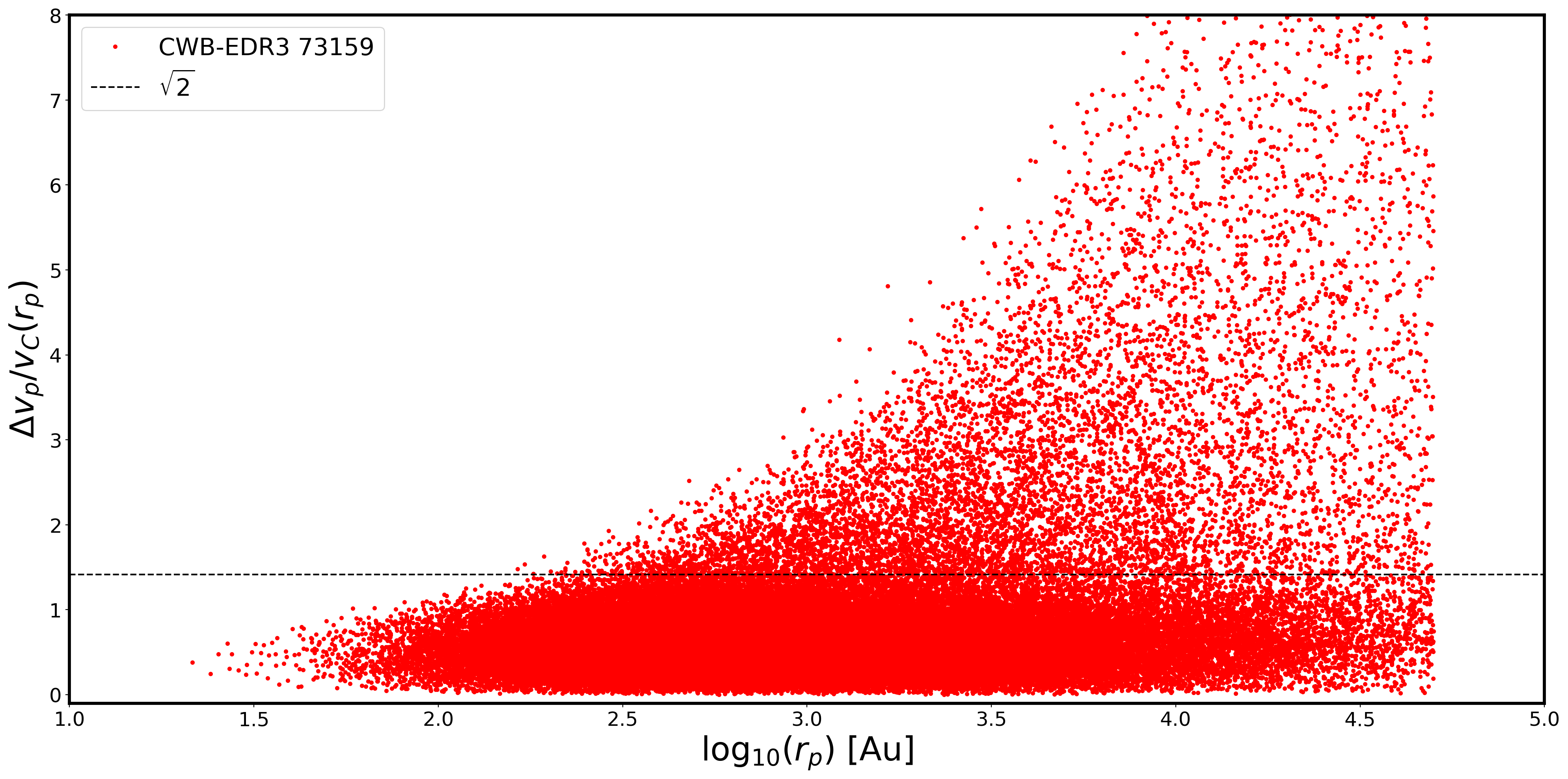} 
			\caption{Scatter plot of projected velocity relative to Newtonian, 
				$\vp/v_c(\rp)$, vs projected separation for CWB-EDR3 sample.  The dashed line at 
				$\sqrt{2}$ indicates the Newtonian limit. The upper cutoff is 
				now slightly fuzzy due to the additional dependence of velocity ratio on mass.  
				Note that in later analysis we only use the sample at $\rp > 5 \kau$ so $\log_{10} r_p > 3.7$, 
				so incompleteness in the upper-left region is not an issue.   }  
			\label{fig:rp_vratio}
		\end{center}
	\end{figure*}

	For the surviving 73,159 candidate binaries, we show
	a plot of projected velocity difference vs projected separation 
	in Figure~\ref{fig:rpvp}; this shows a clear excess approximately
	as expected for bound binaries, with an overdense cloud following
	a locus $\vp \sim 1 \kms (\rp / 1 \kau)^{-0.5}$. 
	We note that our sample starts to miss true binaries at 
	projected separations below  $\rp \la 0.6 \kau$, due to the
	$3 \kms$ velocity threshold, 
	but this $\rp$ is much  smaller than the separations of interest 
	below. At $\rp > 5 \kau$ the threshold includes pairs with velocity 
	difference far above the bound limit, 
	which are interesting for assessing sample contamination as seen below.

	It is more informative to rescale to the typical Newtonian
	orbit velocity, so we next estimate masses for each binary 
	using an estimated mass/luminosity relation:  here,
	we adopt the main-sequence $M_I(mass)$ relation of \textit{Version 2021.03.02} from \citet{Pecaut_2013}, 
	and the $V - I, M_I$ colour relation from the same,
	where $M_I$ denotes absolute magnitude.  From those we
	apply the colour relation given in Table~C2 of \citet{GaiaEDR3_2021} to predict
	$G$ magnitude from $V$ and $I$ magnitudes as 
	\begin{eqnarray} 
		G & \simeq & V \; - \; 0.01597 \; + \; 0.02809 (V-I) \; 
		- \; 0.2483 (V-I)^2 \nonumber \\ 
		& & + 0.03656 (V-I)^3 \; - \; 0.002939 (V-I)^4  
		\label{eq:gvi} 
	\end{eqnarray} 
	to obtain 
	a predicted relationship between absolute GAIA magnitude 
	$M_G$ vs mass; we then fit to this to obtain an approximate
	mass/$M_G$ relation 
	\begin{equation}
		\frac{M}{\msun} = 10^{0.074 (4.69 - M_G )}  
		\label{eq:mass-mg} 
	\end{equation} 
	Then, for each star we have $M_G$ directly from $G$ and 
	parallax distance, 
	hence an estimated mass follows.   Since the luminosity(mass) 
	relation is rather steep, small errors in $G$ or distance
	have relatively little effect on mass estimates below.   

	
	For each candidate binary we then define 
	\begin{equation} 
		v_c(\rp) \equiv \sqrt{G M_{tot} / r_p} 
		\label{eq:vc} 
	\end{equation} 
	as the estimated circular-orbit velocity at the current 
	{\em projected} separation; 
	for each candidate binary, we then divide the measured projected velocity 
	difference by the above to obtain a dimensionless ratio 
	\begin{equation} 
		\vtilde \equiv v_p / v_c(\rp) \ ; 
	\end{equation} 
	a scatter plot of this ratio is shown in 
	Figure~\ref{fig:rp_vratio}, and 
	various histograms of this ratio  are compared with models below.

	\subsection{Transverse velocity errors} 
	
	We have estimated relative-velocity errors assuming
	uncorrelated errors between the two components of the binary, 
	simply from the root-sum-square of the quoted rms errors in $\mu_\alpha$
	and $\mu_\delta$ for each of the two stars in each binary, 
	and multiplying by distance to obtain the transverse-velocity 
	error. (This should be reasonable as long-range correlated errors should 
	mostly cancel between the two stars).

\begin{figure*}
		\begin{center} 
			\includegraphics[width=\linewidth]{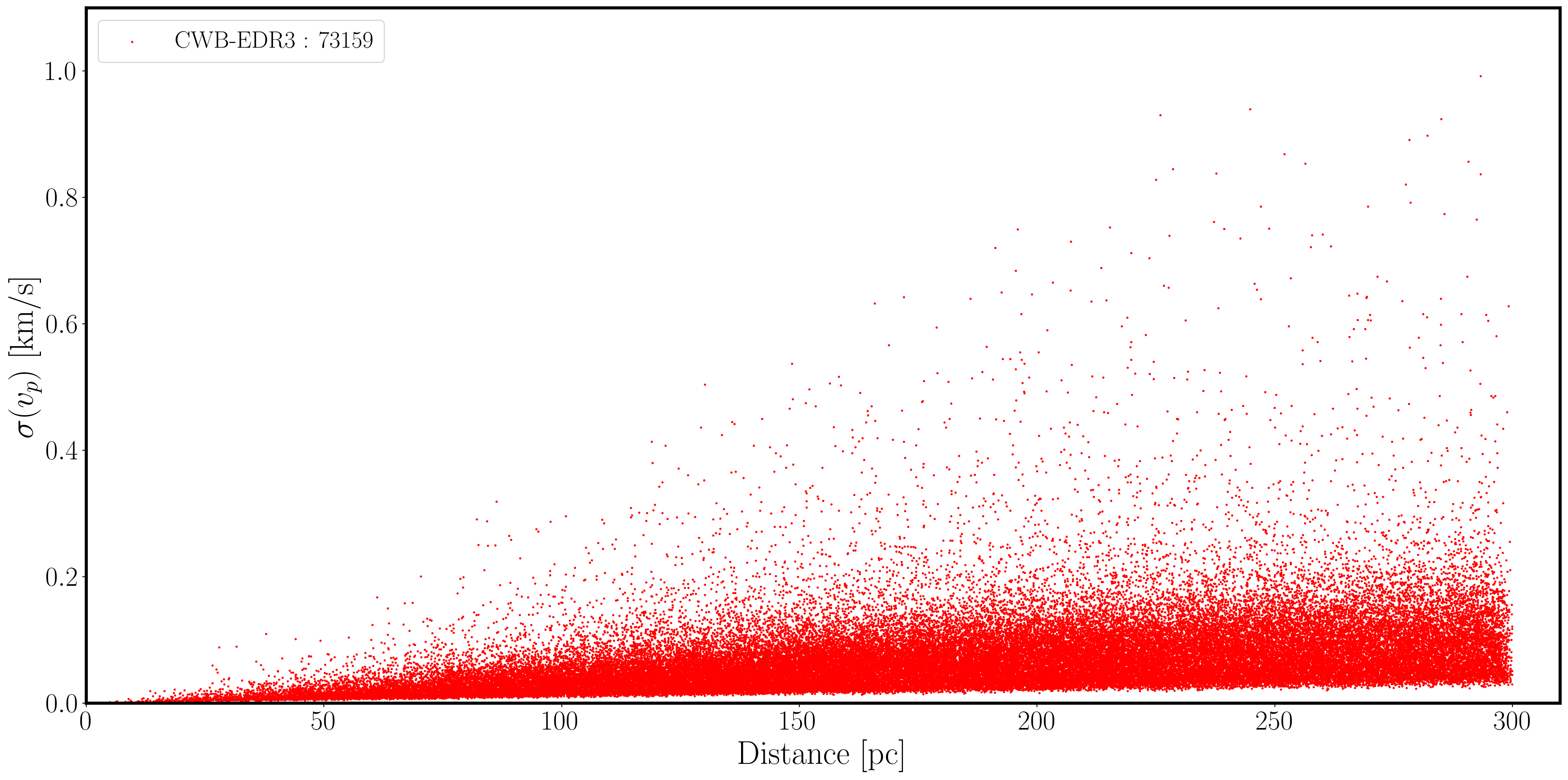} 
			\caption{Scatter plot of rms velocity uncertainty $\sigma(\vp)$ versus
				mean distance for candidate binaries in CWB-EDR3.}
			\label{fig:dsigv} 
		\end{center}
	\end{figure*}

	Table \ref{tab:GDR2ToEDR3TransVelErrors} shows the comparison of the transverse velocity random errors between the CWB-DR2 and CWB-EDR3. We can see with the better quality of data from DR2 to EDR3, and a larger binary sample, the median has decreased to an impressively small value of $\sigma(\vp) \approx 0.06 \kms$.
	We also see a decrease in values when converting to the ratio to circular-orbit velocity, $\sigma(\vp)/v_c(\rp)$, where the median for the full candidate sample has now reduced to 0.06 and the 80th percentile is 0.1; 
	for the ``wide'' subsample with $5 < \rp < 20 \kau$, the
	median is 0.14 and the 80th percentile is 0.26. 
	A scatter plot of $\sigma(\vp)$ 
	versus distance is shown in Figure~\ref{fig:dsigv}; the trend
	with distance is clear, but
	most systems have $\sigma(\vp) \la 0.15 \kms$ even near our
	$300 \pc$ limit.

	\begin{table}
		\caption{Table comparing percentiles of transverse velocity errors 
			and relative to circular velocity,  between CWB-DR2 (PS19) and CWB-EDR3 data sets. 
			Upper table for the full sets, lower table for subsample with $5 < \rp < 20 \kau$. }
		\label{tab:GDR2ToEDR3TransVelErrors}  
		\centering 
		\small
		\begin{tabular}{l l c} 
			\newline
			All $\rp$   &   CWB-DR2   & CWB-EDR3    \\ 
			\hline
			\\
			
			\multicolumn{1}{p{3cm}}{\centering $\sigma(\vp) \  [\kms ]$ \\ (50\%, 80\%, 90\%) \\} & $\approx\,[0.09, 0.14, 0.2]$ &  $\approx\,[0.06, 0.1, 0.13]$   \\
			
			\multicolumn{1}{p{3cm}}{\centering $\sigma(\vp)/v_c(\rp)$ \\ (50\%, 80\%, 90\%) \\} & $\approx\,[0.08, 0.17, 0.26]$ &  $\approx\,[0.06, 0.12, 0.19]$   \\
			\hline
			& & \\
			\multicolumn{1}{p{3cm}}{\centering ($5 < \rp < 20 \kau$)\\} 
			&  CWB-DR2  & CWB-EDR3 \\
			\hline
			\vspace{-0.1cm}
			\\
			\multicolumn{1}{p{3cm}}{\centering $\sigma(\vp)\ [\kms ]$ \\ (50\%, 80\%, 90\%) \\} & $\approx\,[0.09, 0.14, 0.19]$ &  $\approx\,[0.05, 0.09, 0.11]$   \\
			
			\multicolumn{1}{p{3cm}}{\centering $\sigma(\vp)/v_c(\rp)$ \\ (50\%, 80\%, 90\%) \\} & $\approx\,[0.23, 0.39, 0.54]$ &  $\approx\,[0.14, 0.26, 0.34]$   \\
			\hline
		\end{tabular}

	\end{table}

	The latter values are significantly smaller than 1, but not very small, 
	so the effect of random proper motion errors will affect the detailed
	shape of the distributions below.  However, in future GAIA
	data releases these values are expected to reduce by factors
	of at least 2--4 as  proper motion precision scales
	as $\propto t^{-3/2}$ for fixed scan cadence,  so the random errors in proper motions  
	are likely to become relatively unimportant in the medium-term 
	future.  
	
	We note that for a ``typical'' binary below 
	at $\rp \sim 10 \kau$ and $d \sim 180 \pc$, 
	the angular separation is $0.27$ mrad
	or 55 arcsec, so these are very well resolved and the uncertainty
	on $\rp$ is essentially the same as the error on the mean distance,
	typically well below 1 percent and almost negligible.  
	The error on $\vp$ is dominated 
	by random errors on the proper motions, assuming that
	correlated systematic errors mostly cancel between the two components
	of the binary.   Since we are mostly interested in statistical
	distributions, the effect of random errors is modest as long
	as these are not larger than $\sim 0.25$ in $v_p / v_c(\rp)$.  
	Note that for systems with small observed ratios 
	$\vp/v_c(\rp) \sim 0.5$, the {\em fractional} uncertainty in this
	ratio is rather large; 
	however such systems still have a high probability
	of the true ratio being $\la 0.8$, so this scatter is relatively
	unimportant.   For systems with $\vp/v_c(r_p) \ga 1$, the fractional
	uncertainty is relatively modest; though possible non-Gaussian
	errors in the GAIA data remain a concern, this should improve in 
	future GAIA releases as more observing epochs become available
	to reject outliers.

	In summary, we note that the velocity precision for our current sample has improved over the PS19 sample, since
	the improved precision of GAIA EDR3 over DR2 outweighs our expanded distance and magnitude limits.  
	The precision will continue to increase with future DR4 and beyond in the extended mission, 
	so GAIA random errors will become negligible
	compared to other sources of uncertainty (especially contamination from triple systems, see below).

	\subsection{Comparison with El-Badry, Rix and Heintz Wide Binary EDR3 sample} 
	
	Here we note that a catalogue of candidate binaries 
	in GAIA EDR3 has been published by \citep[][hereafter \citetalias{Badry_2021}]{Badry_2021}. 
	Much of our sample selection was completed independently 
	prior to the appearance of \citetalias{Badry_2021}, but here we give a brief comparison. 
	
	\begin{table*}
		\caption{Comparison of Wide Binary selection between our criteria and ERH21  }
		\label{tab:WBEDR3compERH21}  
		\centering 
		\begin{tabular}{l r r r} 
			
			Criteria  &   WB-EDR3  & ERH21     \\ 
			\hline 
			Magnitude $G$  & $<17$ & $\neq {\tt NULL}$    \\
			Parallax $\bar{\omega}\,[mas]$ ($d\,[\pc]$)  & $>\frac{10}{3}\,mas$ ($<300\,\pc$) & $>1\,mas$ ($<1\,\kpc$) \\
			Projected Separation $r_p$  & $\leq50\,kAU$ ($\sim 0.242\,\pc$) & $\leq206.265\,kAU$ ($1\,\pc$)   \\
			Parallax of both members $|\bar{\omega}_1 - \bar{\omega}_2|$ &$\leq4\sqrt{\sigma^2_{\bar{\omega}_1} + \sigma^2_{\bar{\omega}_2} }$ & $ \displaystyle <b\sqrt{\sigma^2_{\bar{\omega}_1} + \sigma^2_{\bar{\omega}_2} } \ 
			\begin{cases}  b=3\, , \, \theta_p\geq\,4\arcsec \\ b=6\, , \,\theta_p<\,4\arcsec  \end{cases}  $ \\ 
			Projected velocity threshold $v_p$  & $\le 3 \kms$ & $\le 2.1 \kms (\rp / 1\kau)^{-0.5}$   \\
			\hline 
		\end{tabular}
		
	\end{table*} 
	
	As seen in Table \ref{tab:WBEDR3compERH21}, \citetalias{Badry_2021} chose  $1 \kpc$ distance limit with Gaia magnitude cuts at $G\neq {\tt NULL}$, while we selected our primary sample at $300 \pc$ with $G<17$.  Note that  
	\citetalias{Badry_2021} do use other cuts on relative errors, 
	which do introduce some implicit magnitude-dependence, as fainter stars are more likely to fail those cuts. 
	Our selection adopts a fixed projected separation of $r_p\leq50\kau $ with parallaxes consistent within 4 times the combined uncertainty $\leq4\sqrt{\sigma^2_{\bar{\omega}_1} + \sigma^2_{\bar{\omega}_2} }$ and a projected velocity
	threshold $v_p \le 3 \kms$ (the same as in \citetalias{Pittordis_2019}). Whereas \citetalias{Badry_2021} use $r_p\leq1\pc$, with parallaxes to both members satisfying \citetalias{Badry_2021} Equation 2 as follows:
	\begin{eqnarray} 
		|\bar{\omega}_1 - \bar{\omega}_2| & \leq  & b\sqrt{\sigma^2_{\bar{\omega}_1} + \sigma^2_{\bar{\omega}_2} }  \nonumber \\
		b = 3 , \,\, & \theta_p \geq 4\arcsec \nonumber \\
		b = 6 , \,\, & \theta_p < 4\arcsec 
		\label{eq:badry21} 
	\end{eqnarray}
	In addition, they apply the same separation-dependent 
	threshold for projected velocities $v_p$ from \citet{Badry_2018}, which translates to $v_p \le 2.1 \kms (\rp / 1\kau)^{-0.5}$ 
	(equivalent to the Newtonian bound limit $\sqrt{2} \,\vc$ for 
	a $2.5 \msun$ system, or just above at 
	$1.83 \,\vc$ for a more typical system with mass  
	$1.5 \msun$).   
	This means that our sample extends to substantially higher
	(unbound) velocity ratios at $\rp \ga 3 \kau$, which turns out to be
	useful below for modelling the tail of high-velocity systems.  
	
	There are additional differences in how we cut for clusters, 
	triples, etc, but these turn out to be relatively less important. We cut out known clusters after removing triples and hierarchical systems as described in sections \ref{sec:skycut}, \ref{sec:triples} and \ref{sec:faintstars}. In \citetalias{Badry_2021}, they begin cleaning by counting the number of phase-space neighbours per source from their primary selection that are brighter than $G=18$, consistent with the size and velocity dispersion of a typical cluster. The neighbours per source are defined satisfying projected separations $r_p\leq 5 \pc$ and proper motions $\Delta \mu \leq 5\kms$, with parallaxes to both sources within 
	$\Delta \bar{\omega}\leq2\sqrt{\sigma^2_{\bar{\omega}_1} + \sigma^2_{\bar{\omega}_2} }$
	Any source (or either component from the binary candidate) that contains more than 30 neighbours is removed. This leads to removing all candidates where either component of the binary is a member in another binary candidate; also removing candidates that might be members of small clusters or moving groups that were no caught in the initial cleaning, performed by applying the same method in counting the number of phase-space neighbours but without the magnitude cut.
	
	We have done a cross-match of our sample to \citetalias{Badry_2021} as follows: we first select the subset of \citetalias{Badry_2021} binaries which could in principle have passed our sky, distance and magnitude criteria, i.e., where both stars have $G < 17$, galactic latitude $\vert b \vert > 15 \deg$, difference in parallax distances $\vert d_1 - d_2 \vert \leq \, 8.0 \pc$, projected separation $\rp < 50 \kau$, distance to both stars $d\leq 300 \pc$, parallaxes of both stars consistent with each other within $4\times$ the combined uncertainty e.g., $ \vert d_1 - d_2 \vert \leq 4\sigma_{d}$, projected velocity difference $\Delta v_p \le 3 \kms$, and applying cuts from \citet{Arenou_2018}. The combination of these cuts gives
	a subsample of 65,234 candidate binaries from \citetalias{Badry_2021} which could 
	in principle pass our other cuts, here called the ERH-cut sample.  Cross-matching those against
	our CWB-EDR3 sample of 73,159 candidates, we find that 64,498
	are in common with ERH-cut, while there are 8,661 candidates in our CWB-EDR3 but not found in ERH-cut, and 752 candidates in ERH-cut are not found in CWB-EDR3.

\begin{figure*}
		\begin{center} 
			\includegraphics[width=\linewidth]{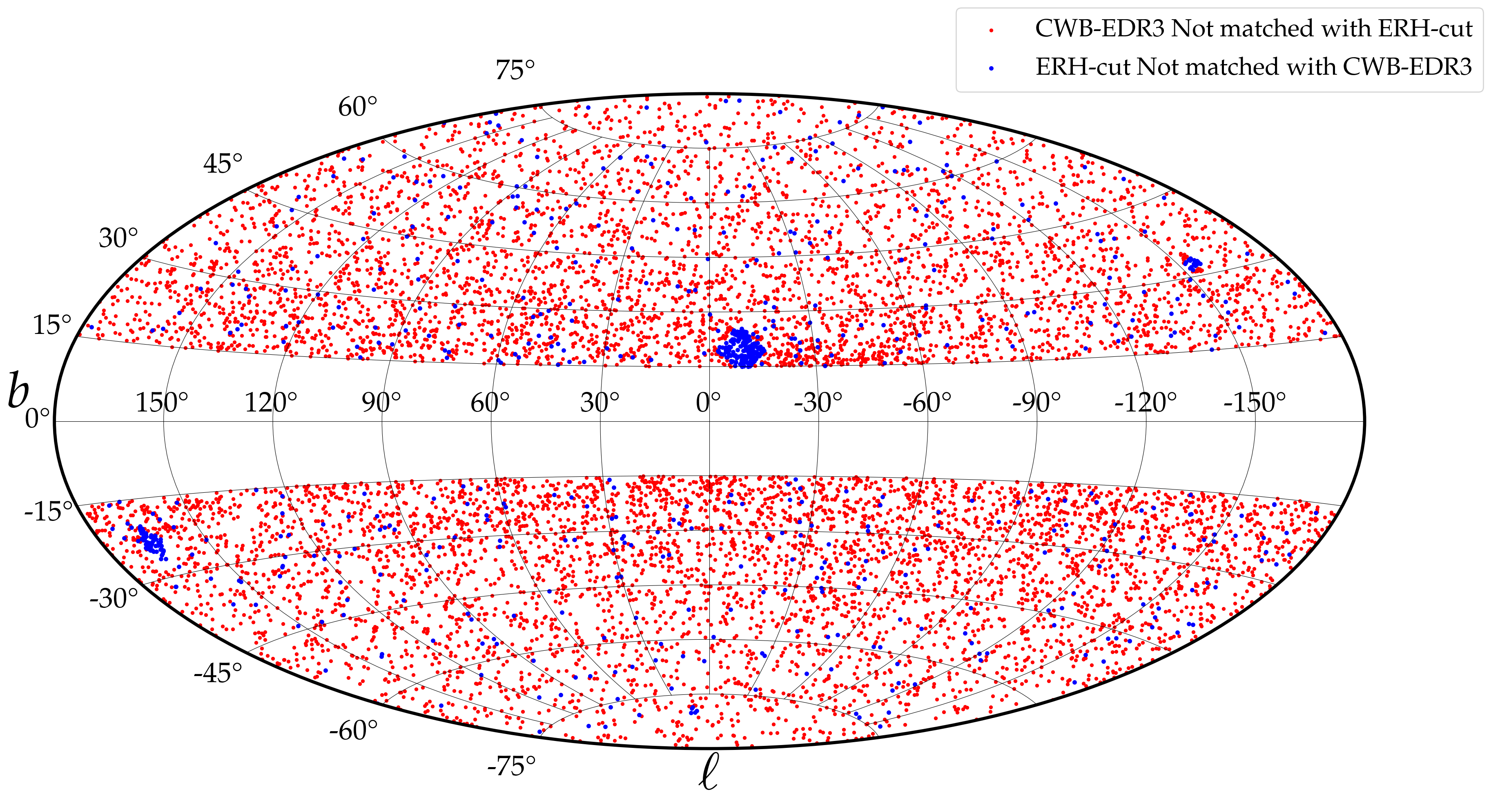} 
			\caption{Aitoff projection of CWB-EDR3 binaries not found in ERH-cut sample (red points), and binaries from ERH-cut  not found in CWB-EDR3 (blue points).}
			\label{fig:ERHaitoff} 
			\includegraphics[width=\linewidth]{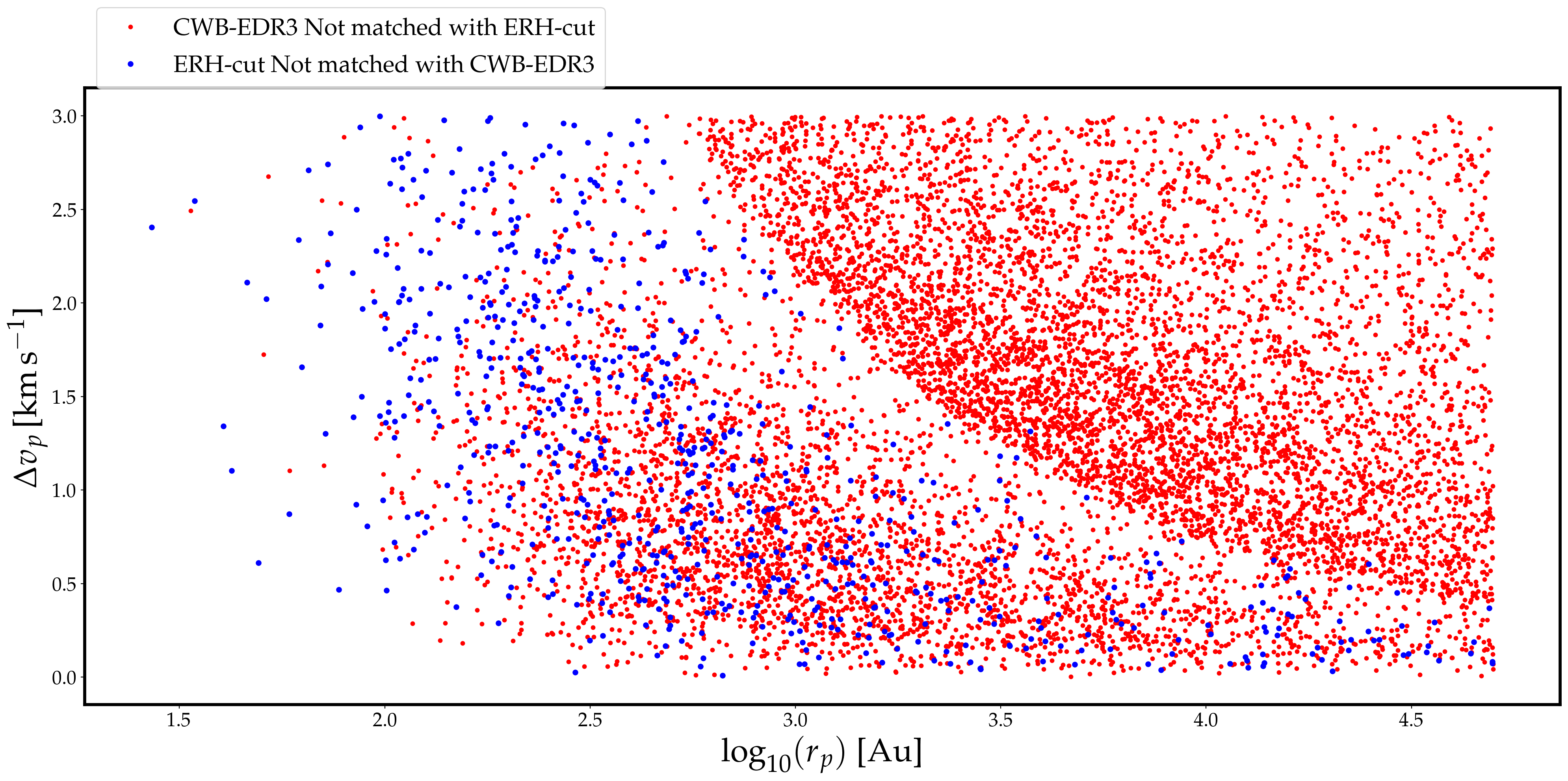} 
			\caption{Scatter plot of projected velocity difference $\Delta\vp$
				vs projected separation for CWB-EDR3 binaries not found in ERH-cut sample (red points), 
				and binaries from ERH-cut not found in CWB-EDR3 (blue points). }
			\label{fig:ERH_VpRp_scatter} 
		\end{center}
	\end{figure*}

	Figures \ref{fig:ERHaitoff} and \ref{fig:ERH_VpRp_scatter} illustrate the comparisons of the 8,661 candidates in our CWB-EDR3 but not found in ERH-cut, and 752 in ERH-cut not found in CWB-EDR3. 
	As seen in Figure \ref{fig:ERH_VpRp_scatter}, most of the additional binaries in our sample are either at larger velocity differences (above the \citetalias{Badry_2021} limit and below our $3 \kms$ limit), or at smaller separations $\rp < 0.5 \kau$ which we do not consider below. 
	
	As seen in Figure~\ref{fig:ERHaitoff}, the binaries in ERH-cut but not in CWB-EDR3 are largely removed 
	due to our additional sky cuts around open clusters.

	In summary we find good agreement between our sample and the ERH-cut sample, 
	and differences in selection criteria have relatively little effect; the additional ``binaries" in our sample are mostly
	potential triples at large velocity differences, which are useful for modelling in later sections.

	\section{Orbit simulations} 
	\label{sec:orbits} 
	
	In this Section we describe the simulations for predicting distributions of projected velocities in Newtonian and MOND models, 
	for various eccentricity distributions, and for binaries, triple systems  
	As noted in Section 3.1 of \citetalias{Pittordis_2019}, we described how velocity ratios are handled within our orbit simulations of observables for the 2D sky-projected velocities as well as 2D projected separations; this is simply because
	radial velocities are not yet available for the large majority
	of our candidate WBs;  though radial velocities may become available in future
	with large spectroscopic surveys 
	such as 4MOST, PFS, WEAVE and MSE, and targeted follow-up of selected systems.

	\subsection{Binary orbits} 
	\label{sec:orbsmg} 
	Similar to Section 3 from \citetalias{Pittordis_2019}, 
	here we simulate a large sample of $\sim 5 \times 10^6$ orbits
	with random values of $a, e$ in both Newtonian gravity and 
	one specific modified-gravity model from \citet{Banik_2018_Centauri} (hereafter \citetalias{Banik_2018_Centauri}). 
	
	We then study the joint distribution of observables, 
	in particular projected separation 
	$r_p$ and relative velocity ratio $\vp / v_c(r_p)$, using several different eccentricity distribution functions.
	
	In the case of modified gravity models, the orbits are generally 
	not closed ellipses, so they are not strictly defined by 
	the standard Keplerian parameters $a, \, e$. As in \citetalias{Pittordis_2018} and \citetalias{Pittordis_2019}, modified-gravity orbits are parametrised using the ``effective'' orbit size $\hat{a}$ and quasi-eccentricity $\hat{e}$ as follows: 
	we define $\hat{a}$ to be the separation at which the simulated relative velocity
	is equal to the circular-orbit velocity (in the current 
	modified-gravity model), then we define $\theta_{\rm circ}$ to be the angle between the relative velocity vector and the tangential direction when the orbital separation crosses
	$\hat{a}$, and then $\hat{e} \equiv \sin \theta_{\rm circ}$;  these definitions 
	coincide with the usual Keplerian $a, e$ in the case of standard gravity. 
	
	{\newa For the eccentricity distributions, we choose 3 example cases: first a flat distribution $f(\hat{e}) = 1$; 
		second the distribution fitted by \citet{Tokovinin_2016}, $f(\hat{e}) = 0.4 + 1.2 \hat{e}$; and thirdly 
		the thermal distribution $f(\hat{e}) = 2\hat{e}$.  We note here some other recent studies:  
		\citet{Tokovinin_2020} and \citet{Hwang_2022} have studied the eccentricity distribution of wide binaries by fitting the  observed  distribution of angles between the sky-projected separation and sky-projected relative velocity; the inferred distributions are clearly rising with $e$.  At large separations, both studies favour a somewhat super-thermal 
		distribution, $f(e) \propto e^{1.35}$; however the difference in predictions between this 
		and the thermal distribution is relatively modest, with median $e$ values $0.7071$ for the thermal 
		distribution versus $0.745$ for $f(e) \propto e^{1.35}$.  
		(We do not use relative angles in the fitting below, though in principle this contains additional information).  } 
	
	
	After integrating these orbits using either of the above
	gravity laws (Newton/GR, Realistic MoND Model with ExFE as described in \citetalias{Banik_2018_Centauri}) and a chosen
	value for external field $g_e \sim 1.2a_0$,  we ``observe'' the
	resulting binaries at many random times and random inclinations to the 
	line-of-sight.  
	
	For each simulated orbit/epoch snapshot, we produce simulated 
	observables including the projected separation $r_p$, projected relative
	velocity $\vp$, and also $\vp/v_C(r_p)$ corresponding
	to our observables from Gaia.

	The radial acceleration law is chosen according to the
	selected gravity theory under consideration with the
	External Field Effect (ExFE) on (see below).  
	For the Newtonian/GR case, we have the standard 
	\begin{equation}
		g_{N} = \frac{G(M_1+M_2)}{r^2}
		\label{eq:g_n}
	\end{equation}
	
	For the MOND-like case with ExFE, 
	we use the fitting function of \citet{McGaugh_2016} (hereafter MLS), 
	sometimes known as the ``radial acceleration relation'', 
	given by 
	\begin{eqnarray} 
		g_{MLS} & = & g_N \, \nu(g_N/a_0) \ ; \nonumber \\ 
		\nu(y) & = & \frac{1}{1 - \exp(-\sqrt{y})} \ ; 
		\label{eq:g_mls} 
	\end{eqnarray} 
	we refer to this $\nu$ function as the MLS interpolating function below. 
	This function is shown by MLS to produce a good fit to rotation curves
	for a large sample of disc galaxies spanning a range of masses;
	it also has the desirable feature that the function $\nu(y)$ converges
	very rapidly to 1 when $y \ga 20$, so deviations on Solar System 
	scales are predicted to be vanishingly small, consistent
	with observational limits.   
	
	
	The only modified-gravity model we consider to compare against our CWB-EDR3 sample is the MOND model with ExFE, using the approximation 
	of \citetalias{Banik_2018_Centauri}; this is given by 
	\begin{eqnarray} 
		g_{N,int} & = & G(M_1 + M_2)/r^2  \\
		g_{N,gal} & = & 1.2 \, a_0 \\ 
		g_{N,tot} & = & \left( g_{N,int}^2 + g_{N,gal}^2 \right)^{1/2} \\ 
		g_{i, EFE} & = & g_{N,int} \nu(g_{N,tot}/a_0) \left(1 + 
		\frac{\kappa(g_{N,tot})}{3} \right) \\  
		\kappa & \equiv & \frac{\partial \ln \nu }{\partial \ln g_{N} } 
		\label{eq:exfe} 
	\end{eqnarray}
	where $g_{N,int}$ is the internal Newtonian acceleration of the binary; 
	$g_{N,gal}$ is the external (Galactic) Newtonian acceleration, 
	$g_{N,tot}$ is the quadrature sum of these, $\nu$ is the MLS function 
	from Equation~(\ref{eq:g_mls})  
	and $g_{i, EFE}$ is our model MOND-ian 
	internal acceleration, approximating the 
	application of the external field effect. 
	(This is not quite an exact solution of the MOND-like equations,
	but is shown by \citetalias{Banik_2018_Centauri} to be a good approximation to the full 
	numerical solution).  
	
	Above,  the observed  Galactic rotation values 
	$v_{LSR} \simeq 232 \kms$ and $R_0 \simeq 8.1 \kpc$ imply 
	a total Galactic acceleration close to $1.75 \, a_0$, hence we 
	require $g_{N,gal} \nu(g_{N,gal}/a_0) \approx 1.75 \, a_0$. Solving this 
	leads to $g_{N,gal} \approx 1.16 \, a_0$ as above and 
	$\nu \approx 1.51$, in reasonably  
	good agreement with the estimated baryonic contribution to the Galactic
	rotation (as expected, since the MLS fitting function was derived by 
	fitting to a sample of external spiral galaxies with well-observed rotation
	curves, so this is consistent with our Galaxy being typical).

	As in \citetalias{Pittordis_2019}, {\newa at wide separations beyond $7 \kau$, 
		this specific modified gravity model produces 
		typical accelerations which saturate at $\approx 1.35 \times$ larger than Newtonian}, 
	and thus binary orbital velocities boosted by $\sim 16 \%$.

	\begin{figure*} 
		\begin{center} 
			\includegraphics[width=\linewidth]{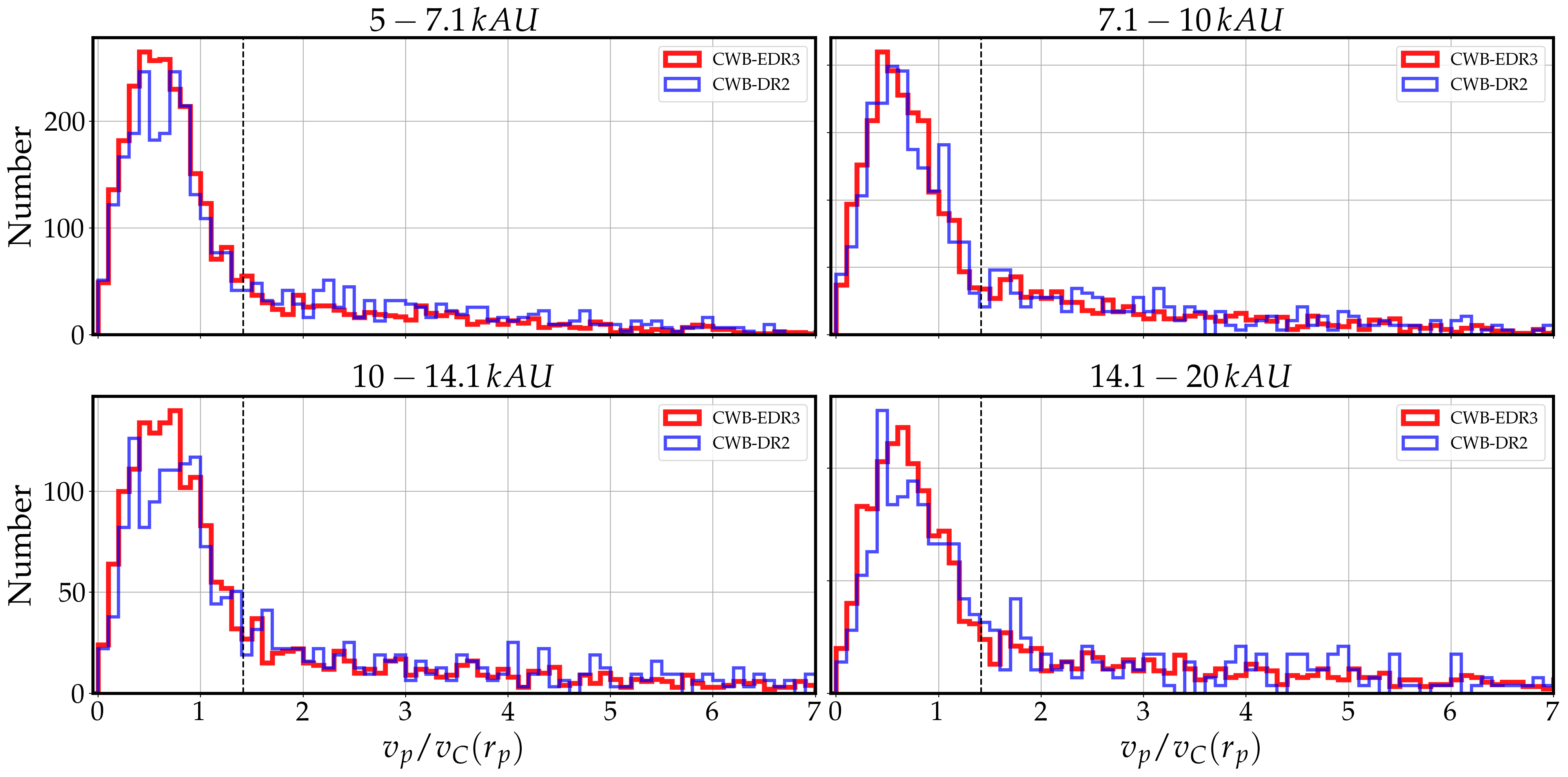} 
			\caption{ 
				Histograms of velocity ratio, $\vratio$, for observed 
				binaries: the four panels show four bins of projected separation as 
				labelled above each panel.  The histograms shows our observed 
				CWB-EDR3 (red) and the CWB-DR2 (blue) sample; numbers in the latter are 
				renormalised to match the total number in the CWB-EDR3 sample.} 
			\label{fig:Gdr2hist} 
		\end{center}
	\end{figure*}
	
	\begin{figure*} 
		\begin{center} 
			\includegraphics[width=\linewidth]{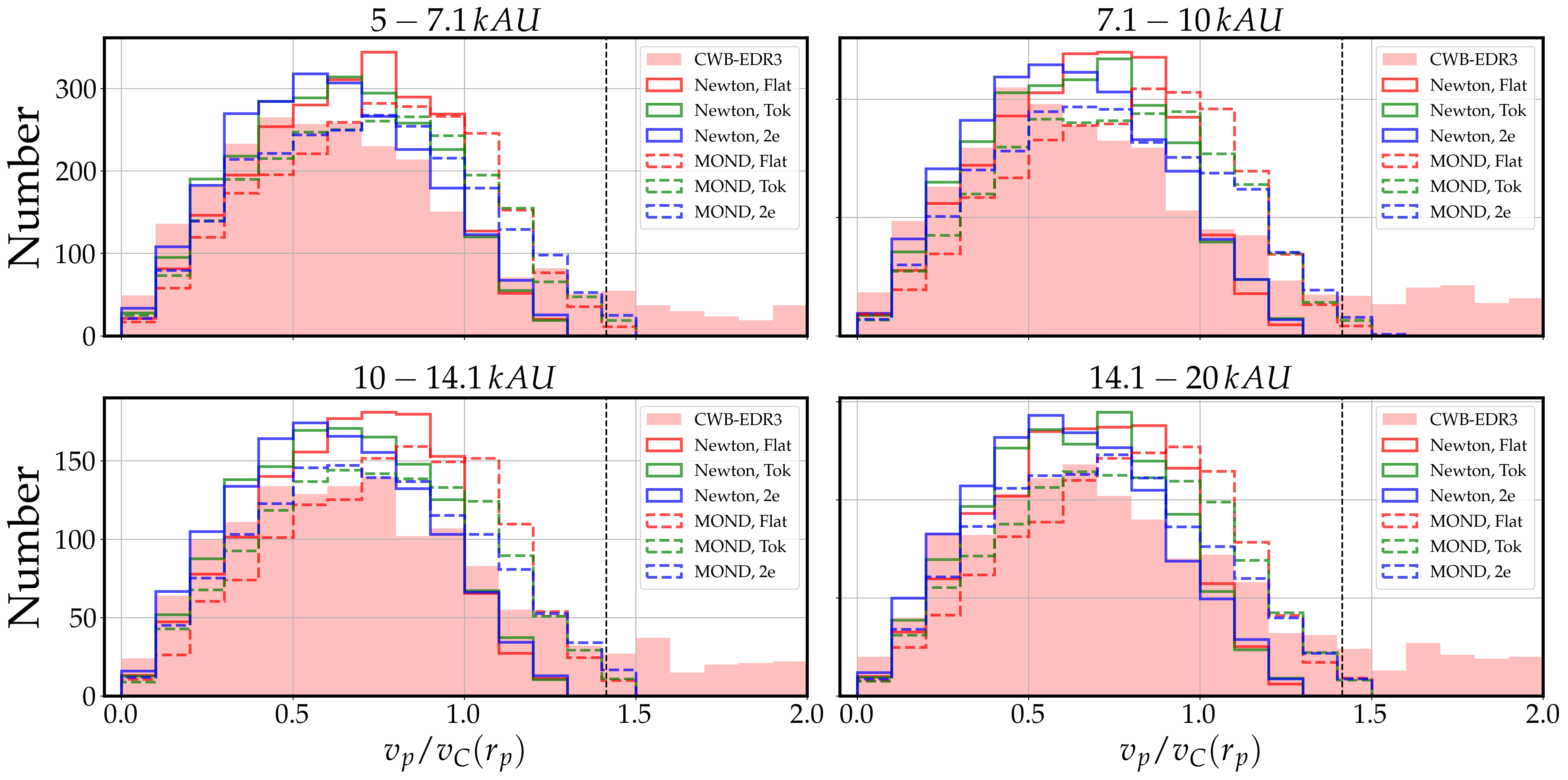} 
			\caption{ Histograms of velocity ratio, $\vratio$, for observed and simulated 
				binaries (simulated binaries are pure 2-star systems excluding triples and flybys): the four panels show four bins of projected separation as 
				labelled in the legend. The shaded pink histogram shows our observed 
				CWB-EDR3 (red); solid lines show Newtonian orbits, and dashed lines for the MOND model. 
				Line colours show eccentricity distribution: flat $f(e) = 1$ (red), Tokovinin (green) and $f(e) = 2e$ (blue). 
				Simulated histograms are normalised to the number of observed systems at ratio $\le 1.6$.    } 
			\label{fig:mghistCut2}
		\end{center}
	\end{figure*} 
	
	In Figure~\ref{fig:mghistCut2}, we show histograms of 
	the ratio $\vratio$ for four bins of projected separation,
	and two gravity acceleration models: standard Newtonian,
	and the Banik$\sim1.2a_0\,$Realistic ExFE from \citetalias{Banik_2018_Centauri}; and a 3-Body Newtonian model compared with the GAIA CWB-EDR3 and CWB-DR2 samples.
	Model histograms are scaled to match the total of the data at 
	velocity ratio $\le \sqrt{2}$. 
	
	On inspection of Figure~\ref{fig:mghistCut2}, several features are notable: 
	\begin{enumerate} 
		\item 
		The observed histograms all show a clear peak at a ratio $\sim 0.6$, in rough agreement with 
		either Newtonian gravity or our MOND model.  The shape of the peak is  sensitive to both the eccentricity distribution
		than the gravity model, with flat $f(e)$ giving a slightly pointed peak, while $f(e) = 2e$ makes the peak
		more rounded and shifted to a slightly lower value.  
		\item Comparing to results from PS19 (not reproduced here), we found that  
		simulated histograms for MOND {\em without} ExFE showed a large and obvious
		shift of the peak to larger velocity ratios $\ga 1$, especially in the 
		wider separation bins where the Newtonian acceleration 
		is well below $a_0$;  
		such a large shift in the peak appears clearly inconsistent with the data, 
		so we confirm the conclusion of PS19 that MOND {\em without} ExFE is firmly excluded (though this  
		was already disfavoured on theoretical grounds).  
		\item
		The simulated histograms for MOND {\em with} ExFE show a distinctly larger fraction of binaries
		at velocity ratio $1.1 - 1.5$, compared to the Newtonian cases; that excess is rather insensitive 
		to the $f(e)$ distribution, increasing	 only moderately with projected separation. 
		So as in PS18, this region is the key discriminant between Newtonian vs MOND models. 
		\item The actual data show a clear peak at $\sim 0.6$ as above, 
		then a fall towards $\sqrt{2}$; however this is followed by 
		a prominent tail which slowly declines to much larger ratios $\ga 5$.  
		Clearly, the presence of this tail  makes it hard to decide a preference 
		between Newtonian or MOND-with-ExFE: a smooth downward extrapolation of
		the tail below $\sqrt{2}$  can  account for a significant fraction of observed systems at 
		velocity ratio $1.1 < \vratio < 1.4$, so 
		the tail is not well understood but
		has a major impact on the statistics.    
		This tail is discussed and modelled in the next Section. 
	\end{enumerate}

	\subsection{Triple system simulations}
	\label{sec:triple-sim} 
	As we can see from the histograms in Figures \ref{fig:Gdr2hist} 
	and \ref{fig:mghistCut2}, the observed CWB-EDR3 sample has a velocity distribution described qualitatively by a ``hump+tail'' shape very similar to \citetalias{Pittordis_2019}.  In \citetalias{Pittordis_2019}, we
	selected random `binaries" from copies of the DR2 catalogue with sky positions randomised by a few degrees;  
	this showed that the``tail" is much too populous to 
	explain as random chance projections of unassociated stars; we verify this for our new sample in \ref{sec:Randoms} below. The next alternative was the possibility of co-natal stars born in the same open cluster, which is currently dissolving therefore, having similar velocities and undergoing chance flybys. Although this provided acceptable $\chi^2$ values, there was a potential inconsistency in that the fitted number of flyby events was decreasing with $r_p$, while simulations predict a rising distribution, due to the increasing phase-space volume at larger projected separation.  It was then suggested by \citet{Clarke_2020} that triple systems with an unseen or unresolved third star could account for the tail, and the simulated distribution was found to produce a good fit to the tail, though the fraction of triples required is quite large (fixed at 50 percent in \citet{Clarke_2020}). 
	Below, we will model the observed distribution with a mixture of binary, triple and flyby systems. 
	
	To simulate the triple systems, we generate two binary orbits made up of three stars. We label star 1 as the single star, and stars 2 \& 3 as the inner binary, so the outer orbit is star 1 orbiting the barycentre of stars 2 and 3. 
	We choose random masses for the three stars with a flat distribution up to $0.7 \msun$, then declining
	as $m^{-2.35}$; we pick three random masses, relabel so that $M_2 > M_3$, and apply the constraint that stars 1 and 2 are $\ge 0.5 \msun$,  while star 3 is $\ge 0.01 \msun$ and may be unobservable.  
	
	For orbit sizes, we choose the outer orbit size $\hat{a}_{out}$ from a flat distribution in $\log_{10} (\hat{a}_{out} / 1 \kau) \in (0,2)$, and the inner orbit size $\hat{a}_{inn}$ is chosen from a flat distribution in $\log_{10} \hat{a}_{inn}$, with a minimum at 0.1 AU and a maximum at $0.3 \times$ the outer orbit; this is an approximate  criterion for long-term stability. 
	
	We generate eccentricities from one of three distributions: either flat $f(e) = 1$, an intermediate case
	from \citet{Tokovinin_2016} (hereafter Tokovinin), and $f(e) = 2e$. 
	Both inner and outer eccentricities use the same distribution function, but drawn independently of each other. 
	
	For simplicity, we treat the inner and outer orbits as independent.   
	We solve for the two orbits independently in their own planes, and then apply a random 3D rotation matrix,
	$\mathbf{R}$,  to the inner orbit relative velocity to generate a random relative alignment between the two orbits. 
	Next the system is ``observed" at 10 random phases and 10 random viewing directions for each phase. At each simulated
	``observation" we evaluate the projected separation, and the 3D velocity difference between star 1 and 
	the ``observable center" of stars 2+3, as follows. 
	\begin{equation}
		\mathbf{v}_{3D,obs} = \mathbf{v}_{out} - f_{pb} \, \mathbf{R}\mathbf{v}_{inn}
	\end{equation}
	where $\mathbf{v}_{out}$ is the outer orbit velocity (star 1 relative to the barycentre of 2+3), and $\mathbf{v}_{inn}$ is
	the relative velocity between stars 2+3.   The latter velocity is rotated and scaled down by a dimensionless factor
	$f_{pb} \le 0.5$, defined as the fractional offset between the ``observable centre" and the barycentre of stars 2+3, 
	relative to their actual separation.  
	The ``observable centre" is defined according to the angular separation: if less than 1 arcsec, we assume stars 2+3 are
	detected by GAIA as a single unresolved object, and use the luminosity-weighted centroid (photocentre) of the two. 
	Otherwise for separation $> 1\,$arcsec, we assume stars 2 and 3 are detected as separate objects, or star 3 is unobservably faint, and we take the position of star 2 alone as the observable centre.  Therefore, this results in a fractional offset 
	\begin{equation} 
		f_{pb} = \begin{cases} 
			\frac{M_3}{M_2+M_3} - \frac{L_3}{L_2 + L_3}  \ (\theta < 1 \text{ arcsec}  ) \\ 
			\frac{M_3}{M_2+M_3}   \ (\theta \ge 1 \text{ arcsec} ) 
		\end{cases} 
	\end{equation}  
	where the $L_{2,3}$ are the model luminosities. 
	
	The 3D velocity above is then converted to 2D projected velocity according to the random viewing direction, and observables are saved to create simulated histograms for triple systems; this procedure is repeated for both of the gravity models (Newtonian / MOND) and each of the three eccentricity distributions. These simulated triple results are then used in the fitting procedure below.  
	
	\subsection{Random Samples}
	\label{sec:Randoms}
	Using the same method as described in Section 3 of \citetalias{Pittordis_2019}, we estimated the level of contamination of our CWB-EDR3 sample by random chance projections of unrelated stars which just happen to contain small velocity differences. We constructed several randomised samples by starting with our single-star PEDR3 sample, removing one star from each binary, then randomising the RA/Dec values individually by a few degrees (e.g., randomising within $ \in \pm 3 \deg$);  these position shifts preserve global distributions with respect to galactic coordinates, but large enough to eliminate most truly associated stars; we then re-ran the binary search on the position-randomised samples. The number of random ``binaries" was small in each single run, so we repeated 9 times with different random seeds and averaged.  
	
\begin{figure*} 
		\begin{center} 
			\includegraphics[width=\linewidth]{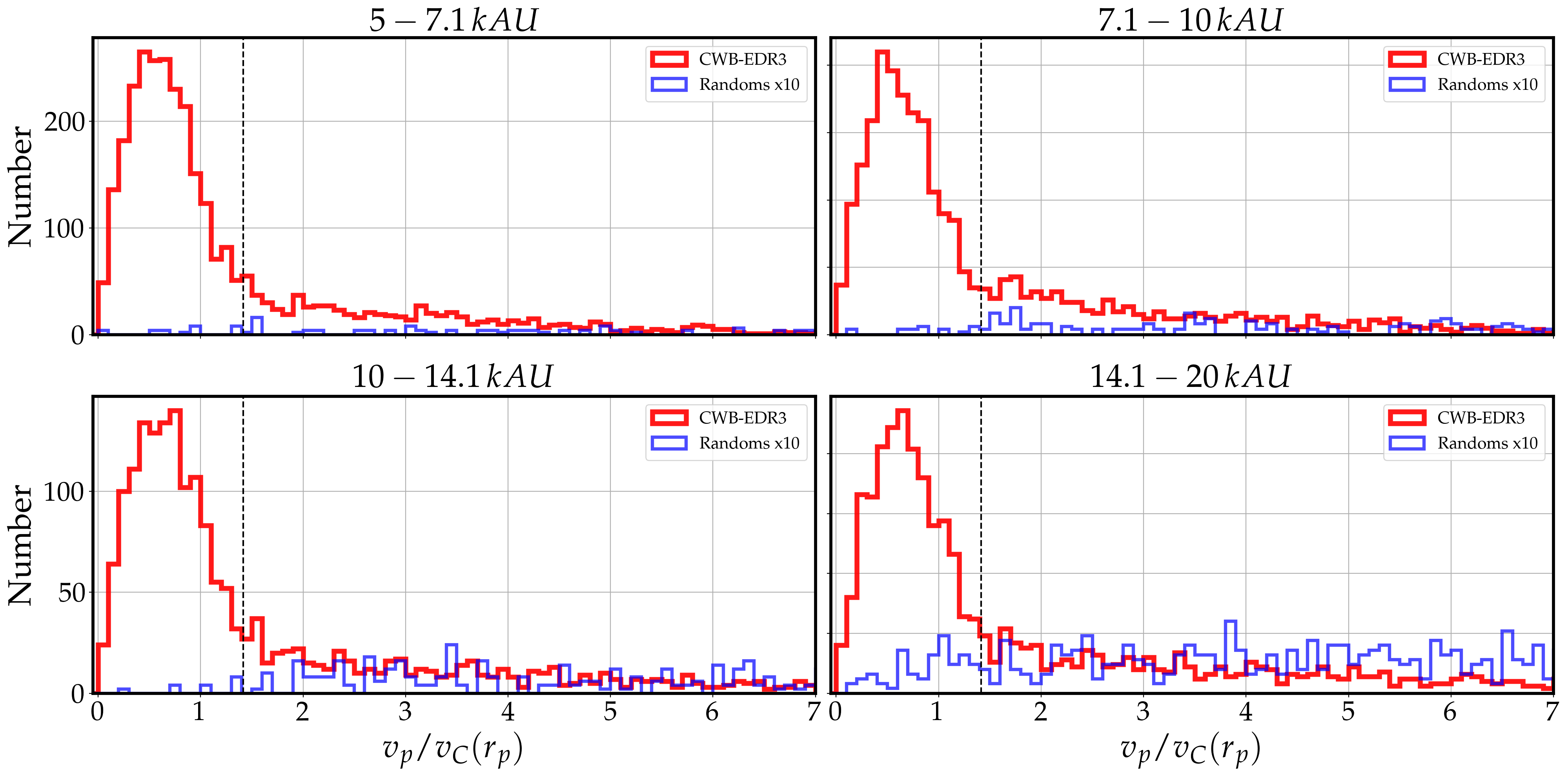} 
			\caption{ 
				Histograms of velocity ratio $\vratio$ for observed and randomised sample. The four panels show four bins of projected separation as titled. The histograms show our observed CWB-EDR3 (red), and the mean of 9 randomised samples (blue), with the mean random sample artificially scaled-up by $10 \times$ to enhance visibility.  } 
			\label{fig:histRands}
		\end{center}
	\end{figure*}

	We show an example histogram for a set of randomised samples compared to our observed CWB-EDR3 data in Figure \ref{fig:histRands}; here the mean of random samples is multipled by $10\times$ for visibility, since the number of randoms is much smaller than the observed data and difficult to see in the histogram.
	
	The main result from our randomised samples is that they contain far fewer candidate ``binaries'' than the observed CWB-EDR3, hence observed data greatly outnumber the randoms. As we can see, the randomised sample show no peak at small velocity ratios, but a fairly smooth distribution with a gradual rise towards larger velocity ratios. We get the same result as
	\citetalias{Pittordis_2019} that the ``tail'' distribution of the observed data is much more populous than our randomised samples, hence the ``tail'' cannot be due to chance projections of unrelated stars.  
	
	\subsection{Flyby simulations} 
	\label{sec:flyby} 
	We find below that fitting purely with binary+triple systems gives reasonable fits, but requires a somewhat improbable 
	result that the ratio of triples to binaries is steeply increasing towards larger projected separations. As in \citetalias{Pittordis_2019},  we therefore
	model an additional population of ``flyby" systems which have small velocity differences (e.g. co-natal stars born in the same birth cluster) on unbound hyperbolic flyby trajectories. 
	
	Similar to \citetalias{Pittordis_2019}, 
	our randomised-position samples show that
	the tail is much too populous to be explained by 
	random chance projections of unrelated stars;  
	one plausible explanation appears to be pairs of co-natal stars
	born in the same open cluster, which therefore have similar
	velocities and are currently undergoing a chance close flyby. 
	Evidence for a population of such ``cold streams'' has been
	given by \citet{Oh_2017}. 
	
	In this case, we would expect two of the three velocity components
	(perpendicular to the escape direction) 
	must be similar in order to get a close flyby, while the velocity
	difference in the escape direction should approximately reflect
	the distribution in ejection velocities from the cluster
	and the time difference between the two ejections. 
	
	During a flyby, the relative velocity will speed up 
	according to a hyperbolic flyby orbit.  To simulate this, 
	we generated random flyby encounters as follows: 
	\begin{enumerate} 
		\item We chose a distribution of impact parameter $b$ 
		with $dn/db \propto b$ up to a maximum value of $300 \kau$ or 1.45 pc. 
		\item We chose a distribution of asymptotic velocity difference 
		$\vinf$  uniform up to $2 \kms$. 
		\item Given random values of $b, \vinf$ generated 
		as above, each such pair produces one flyby encounter.  
		We sampled each hyperbolic flyby 
		at random times while the 3D separation was $\le 300 \kau$,
		and computed the 3D relative separation and relative velocity vectors. We 
		``observed'' these from random viewing angles to produce $\vp, \rp$.  
		\item We truncated the sample as for the DR2 data, 
		at $\rp < 40 \kau$ and  $\vp \le 3 \kms$, 
		and computed the resulting velocity ratio $\vp / v_c(\rp)$. 
	\end{enumerate} 
	For hyperbolic flyby encounters the 3D velocity ratio $v_{3D}/v_c(r)$ 
	is always  $\ge \sqrt{2}$, with the simulations showing a modest  
	pile-up in the distribution just above this value; this pile-up 
	arises from flybys with eccentricity
	not much larger than 1, which speed-up substantially but have
	velocity ratios decreasing towards $\sqrt{2}$ as 
	they approach pericenter.   
	The projection to 2D separation and velocity smears this distribution towards lower ratios, 
	and therefore fills-in the
	gap below $\sqrt{2}$; the result is that simulated fly-bys with a flat 
	distribution of $\vinf$ produce a smooth 
	maximum in the distribution at a ratio below 1.0, 
	with a gently declining  tail at larger ratios, as shown in the next section.      
	\vfill  
	
	\section{Data vs model comparisons}
	\label{sec:comp} 
	
	In this section we fit the observed distributions of $\vp/\vc(r_p)$ in our EDR3 wide-binary sample 
	as a mixture of binary, triple and flyby systems, where the velocity distribution for each is given by the results of
	the simulations in Section~\ref{sec:orbits}.  (Thus, this is an generalised case between \citet{Clarke_2020} who fitted binaries+triples, and \citetalias{Pittordis_2019} which fitted binaries + flybys. )  
	
	{\newa Here, we have applied an additional cut to the sample based on the ``Renormalised Unit Weight Error" 
		or {\tt ruwe} parameter defined in GAIA EDR3; this is a measure of scatter of individual GAIA observations around the 
		basic 5-parameter fit parallax + uniform proper motion, scaled so the median {\tt ruwe} is close to 1.  Objects
		with a {\tt ruwe} value significantly larger than 1.4 are indicative of excess scatter which may indicate a poor fit 
		or a marginally-resolved close pair.  Therefore,  we apply an additional cut that both stars in a candidate 
		binary are required to have {\tt ruwe} $ < 1.4$.   This reduces the wide-binary sample ($5 \kau \le r_p \le 20 \kau$)
		from 9063 systems to 7276 systems,  but significantly reduces the fraction in the high-velocity tail at $\vtilde > 1.4$; 
		of the 1787 pairs rejected by the {\tt ruwe} cut, 68\% had $\vtilde > 1.4$; while of the 7276 surviving systems, only 24\% have  $\vtilde > 1.4$.  } 
	
	We take the four bins as above in projected separation, and fit independently to the observed $\vtilde$ distribution in each separation bin.  
	
	As noted earlier, we found that fitting with only binary+triple systems can produce reasonable fits to the data
	(in agreement with \citet{Clarke_2020}),  but has
	a problem in that the fitted ratio of triple to binary systems is steeply increasing with $r_p$; the 
	fitted {\newa population ratio (number of triples)/(number of pure 2-star binaries) is 1.25  } in the 
	smallest $r_p$ bin, but increases to around 2.75 in the widest $r_p$ bin; this is because the ``tail" becomes
	relatively more populous relative to the ``hump" with increasing $r_p$, and the tail is completely fitted by triples
	since there are no binaries above ratio $1.41$ in Newtonian gravity, or 1.6 in MOND. 
	We therefore modified the fitting to 
	include an additional population of ``flyby" systems as in Section~\ref{sec:flyby}, 
	which naturally tends to increase with $r_p$. 
	
	This leads to an issue in that the triple and flyby populations are somewhat degenerate with each other since they
	both produce similar shapes in the ``tail", so we then fixed the triple/binary population ratio at a constant value;
	we explored fits with ratio of (triple systems)/ pure 2-star binaries) set to 0.5, 0.8 or 1.0, and found that 1.0 is marginally preferred, so
	results below are for this value. 
	We then repeat the fitting for each separation bin with two gravity models and three 
	eccentricity distributions, for a total of $2 \times 3 \times 4$ independent fits.   
	Each fit has only two adjustable parameters,
	the total number of pure-binary systems and the total number of flyby systems; 
	we hold the ratio of triple to pure-binary systems fixed at a selected constant value, here 1.0.  The shape of 
	each of those is constrained at the output of the simulated orbits, for the corresponding gravity model and 
	eccentricity distribution. We then simply fit the number in each population 
	to minimise the $\chi^2$ residuals between binary+triple+flyby  
	model and the data. 
	
	The fitting range is $0 \le \vtilde \le (5,6,7,7)$ respectively in the four $r_p$ bins; these upper limits 
	are chosen because our $3 \kms$ velocity
	cutoff  leads to some incompleteness for higher-mass binaries at smaller $r_p \sim 5 \kau$ and high $\vtilde \ge 5$. At our 
	90th percentile system mass of $2.02 \msun$, $3 \kms $ maps to velocity ratio $\vtilde = 5.0 \sqrt(r_p/5 \kau)$ 
	so our $3 \kms$ threshold includes nearly all systems up to the above $\vtilde$ limits.

	The results of the fitting procedure are shown in Figures~\ref{fig:3BodyNewtonFlat} - \ref{fig:3BodyBanik2Ecc}  below.
	Here Figures~\ref{fig:3BodyNewtonFlat} - \ref{fig:3BodyNewton2Ecc} show the fits with Newtonian
	gravity and the three simulated eccentricity distributions (flat, Tokovinin, $2e$) respectively,  
	and Figures~\ref{fig:3BodyBanikFlat} - \ref{fig:3BodyBanik2Ecc} show the
	corresponding results for the  specific modified-gravity model described in Sec.~\ref{sec:orbsmg},
	and the same three eccentricity distributions.  
	
	Some selected values from the fitting results are listed in Tables~\ref{tab:3BodyNewtonFlat} - \ref{tab:3BodyBanik2Ecc}; 
	for each separation bin and model, this lists the observed numbers of
	EDR3 systems and fitted number of binary, triple and flyby systems, in three selected ranges of $\tilde{v}:$ 
	firstly $ 0 \le \tilde{v} < 7$, secondly $0 \le \tilde{v} \le 1.4$, and also $1.1 \le \tilde{v} \le 1.4$, 
	with the latter range being particularly sensitive to modified-gravity effects.

	\begin{figure*} 
		\begin{center} 
			\includegraphics[width=0.8\linewidth]{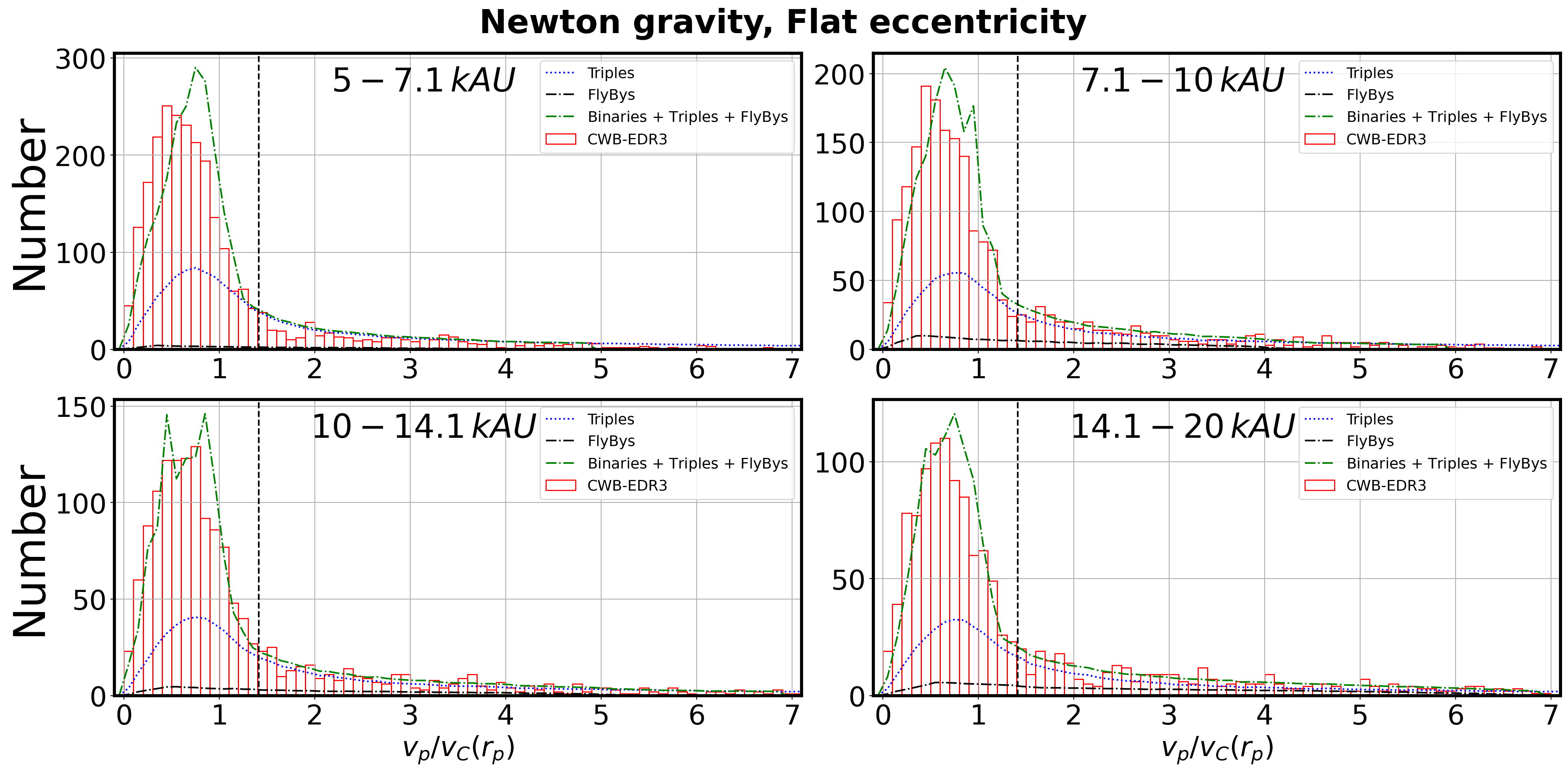} 
			\caption{ Fits of velocity ratios using the Binaries+Triples+FlyBys model for Newtonian gravity and
				all orbits drawn from a flat eccentricity distribution.  The four panels show four bins of projected separation as labelled.
				The red histogram shows the CWB-EDR3 data. The blue-dotted line is the fitted triple population; black half-dashed line are the FlyBys; green dashed line is the total, so the difference is fitted true binaries.
				\label{fig:3BodyNewtonFlat}   } 
			\includegraphics[width=0.8\linewidth]{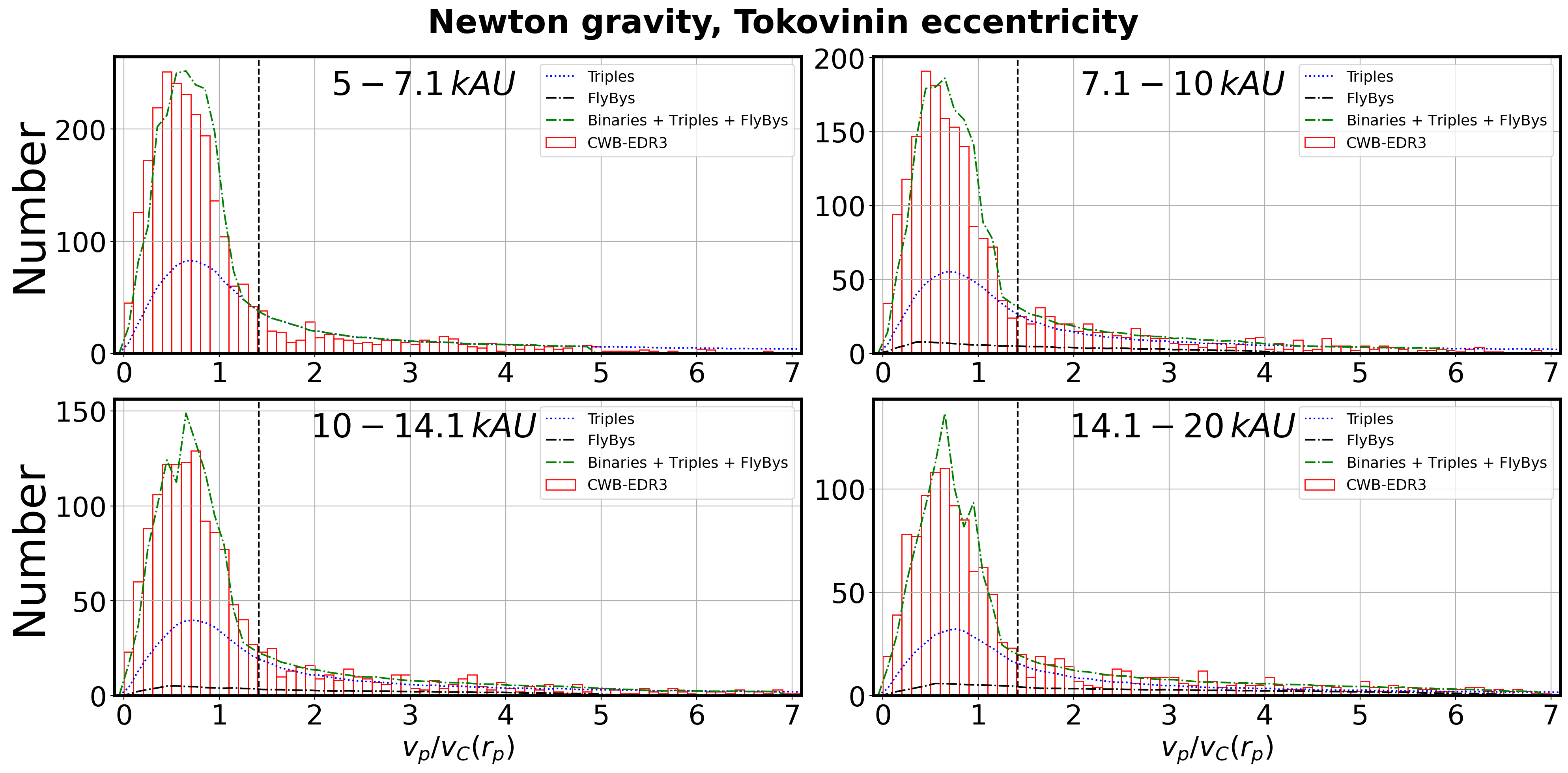} 
			\caption{ Same as Figure~\ref{fig:3BodyNewtonFlat}, except that model orbits are drawn from a Tokovinin eccentricity distribution. \label{fig:3BodyNewtonTok} } 
		\end{center}
		\begin{center} 
			\includegraphics[width=0.8\linewidth]{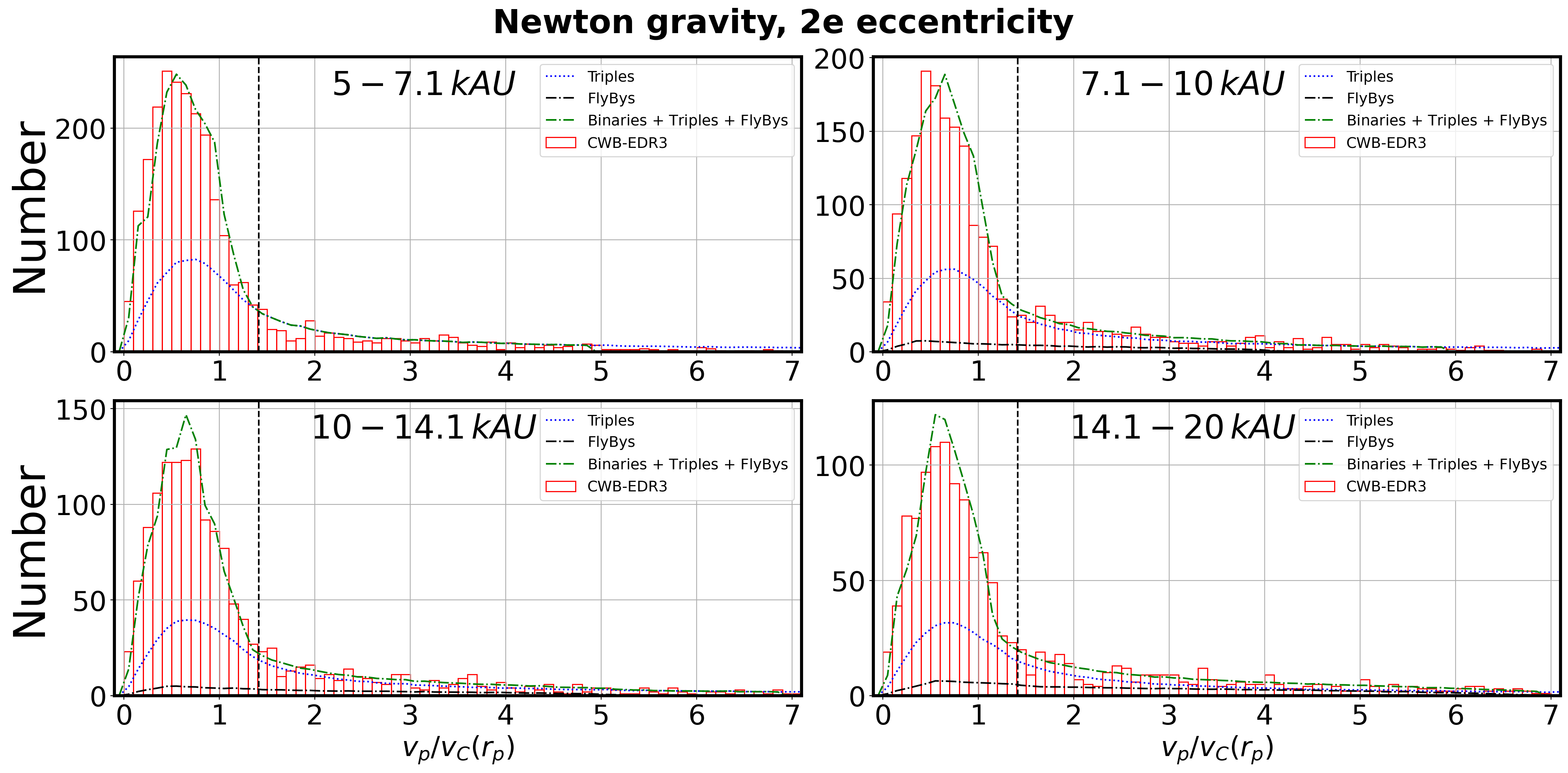} 
			\caption{Same as Figure~\ref{fig:3BodyNewtonFlat} with Newtonian gravity, but model orbits are drawn from a $f(e) = 2e$ eccentricity distribution.  \label{fig:3BodyNewton2Ecc}  } 
		\end{center} 
	\end{figure*} 
	
	\begin{figure*}  
		\begin{center} 
			\includegraphics[width=0.8\linewidth]{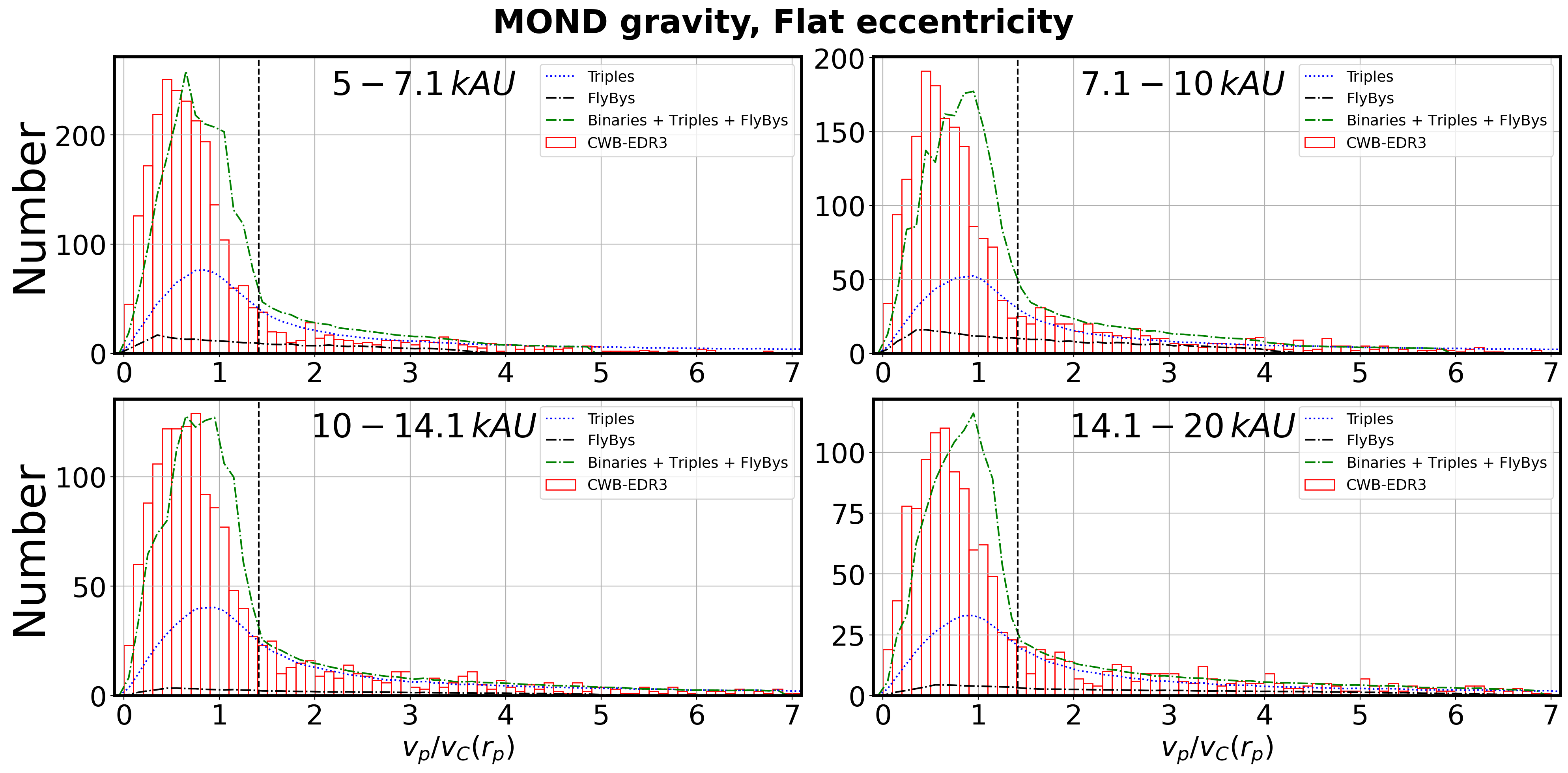} 
			\caption{ Same as Figure~\ref{fig:3BodyNewtonFlat}, but using the realistic MOND gravity model of Sec.~\ref{sec:orbsmg} and all orbits drawn from a flat eccentricity distribution.   
			} 
			\label{fig:3BodyBanikFlat}
		\end{center}
		\begin{center} 
			\includegraphics[width=0.8\linewidth]{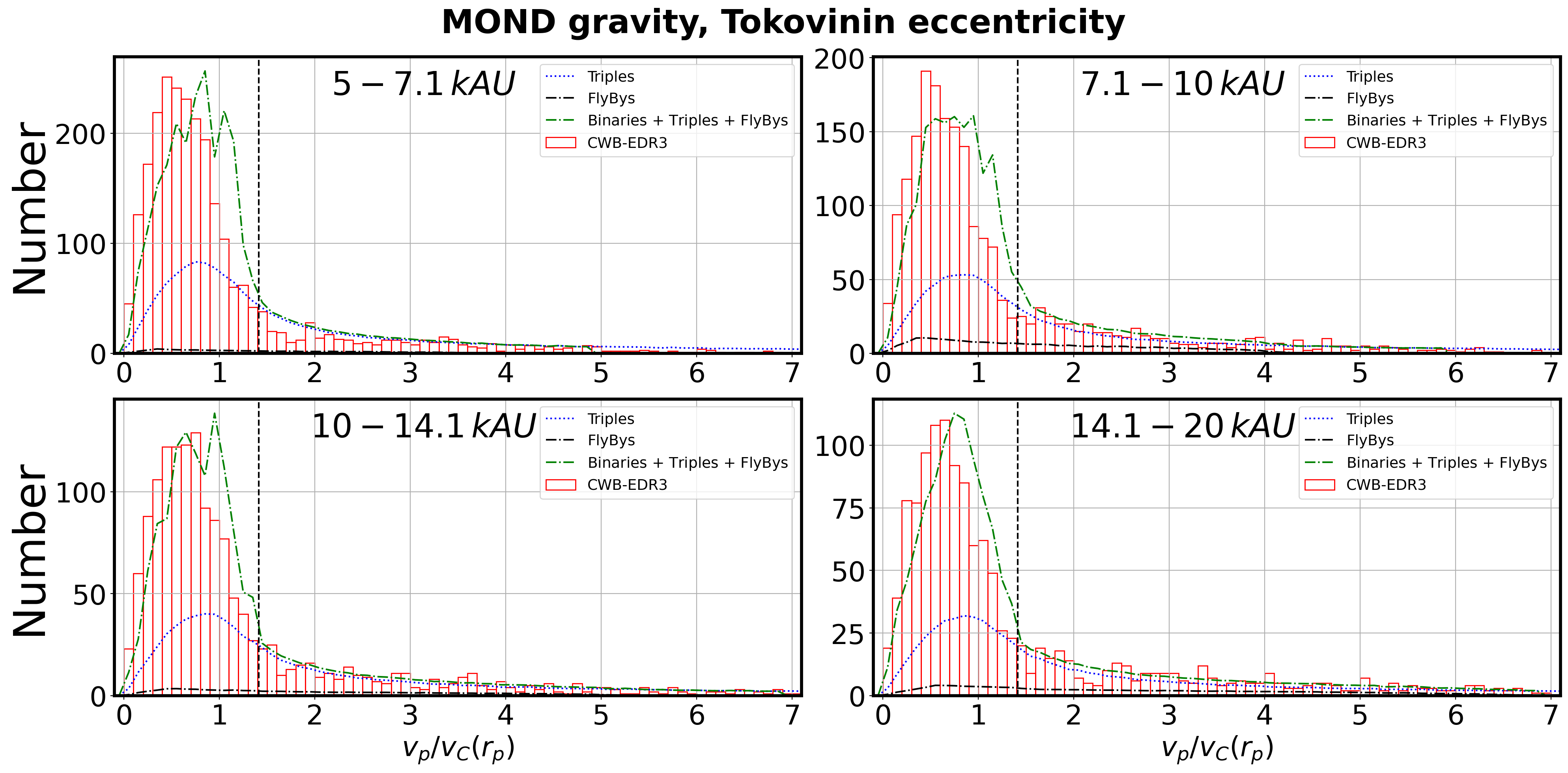} 
			\caption{ Same as Figure~\ref{fig:3BodyBanikFlat}, with realistic MOND gravity model and model orbits drawn from a Tokovinin eccentricity distribution. 
			} 
			\label{fig:3BodyBanikTok}
			\includegraphics[width=0.8\linewidth]{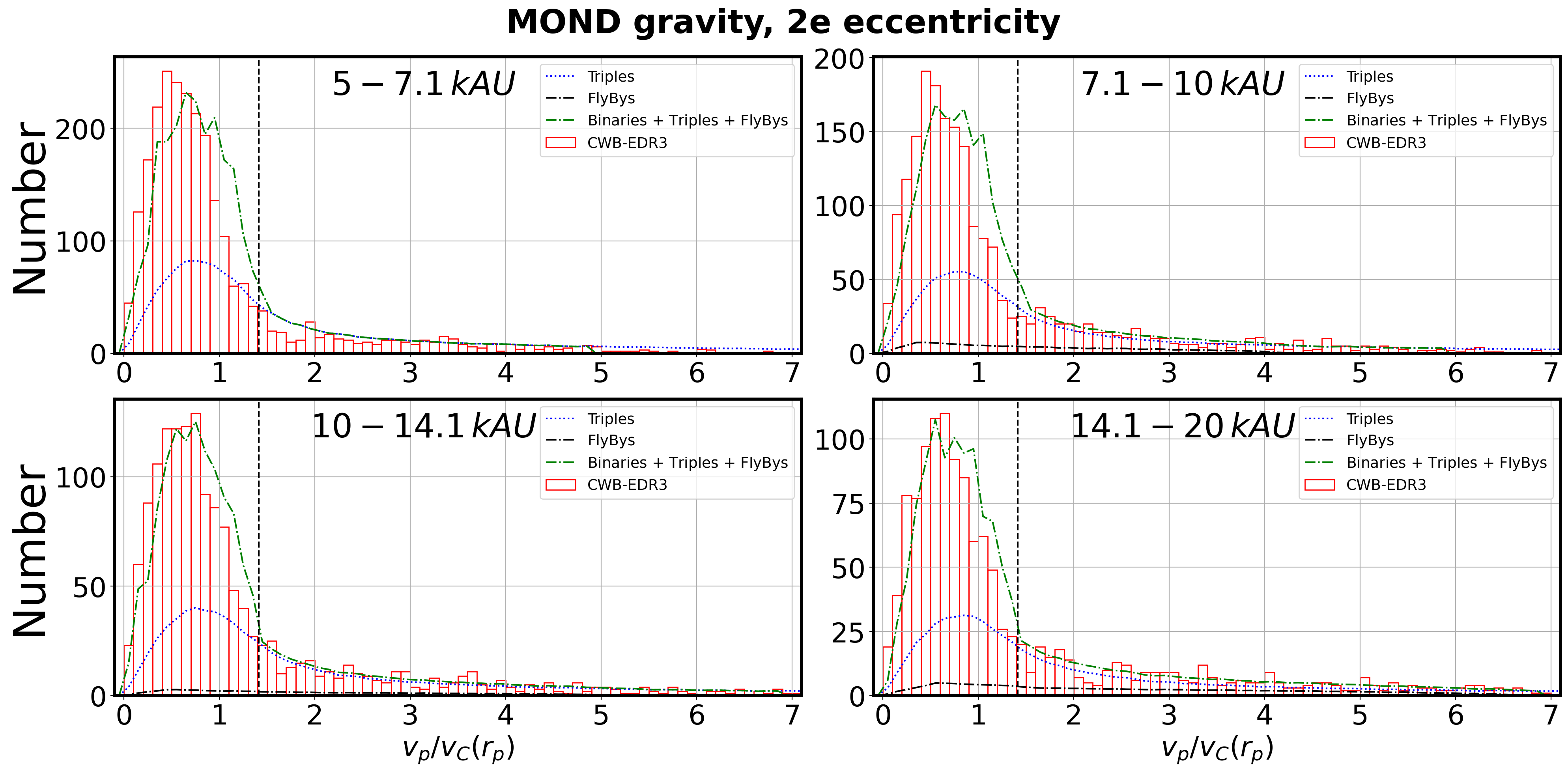} 
			\caption{ Same as Figure~\ref{fig:3BodyBanikFlat}, with realistic MOND gravity model and model orbits drawn from
				an $f(e) = 2e$ distribution.
			} 
			\label{fig:3BodyBanik2Ecc}
		\end{center}
	\end{figure*}
	
	\begin{figure*} 
		\begin{center} 
			\includegraphics[width=\linewidth]{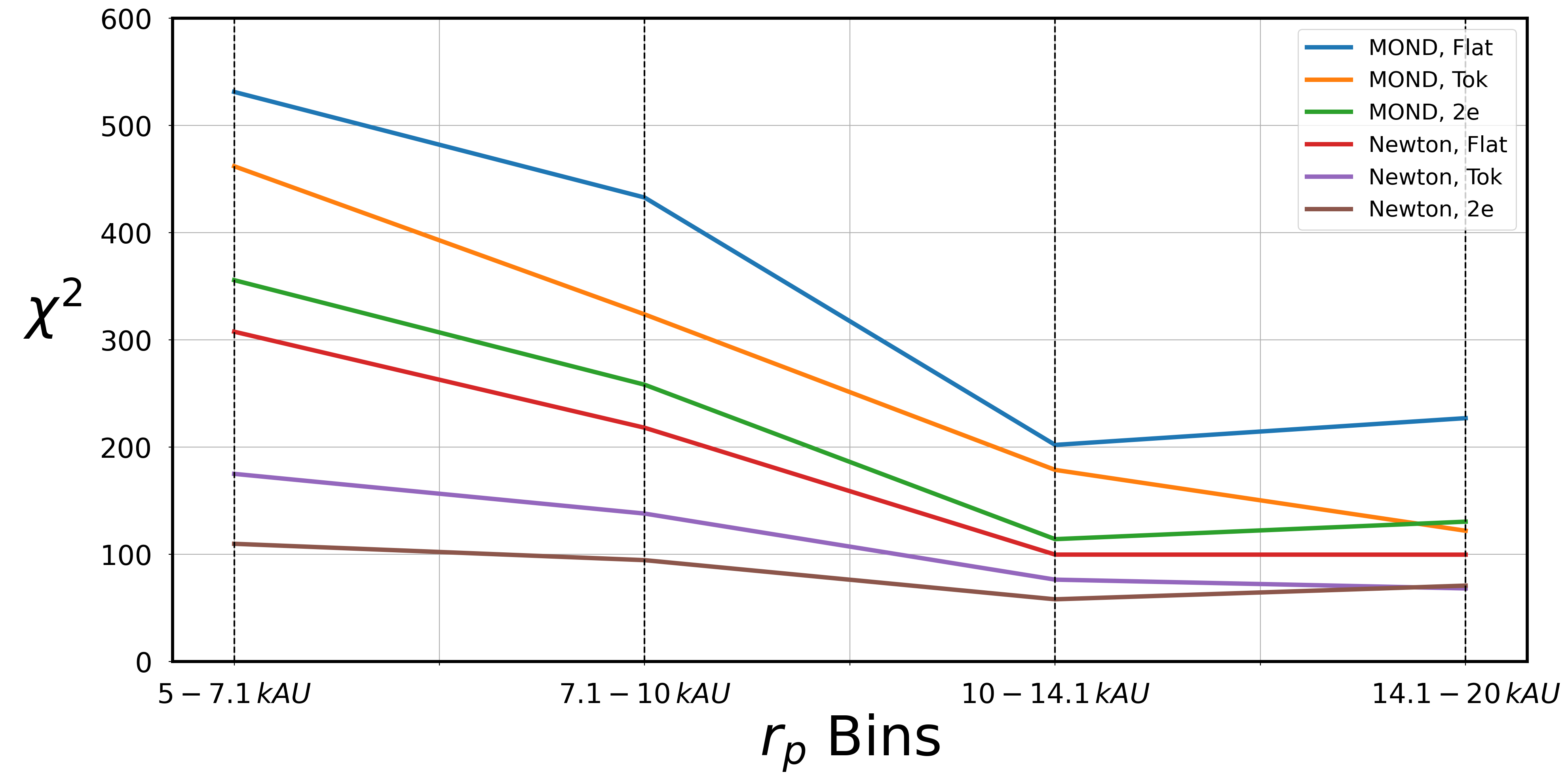} 
			\caption{ Plot comparing the $\chi^2$ fitting results for each of the six orbit models (lines labelled in legend) against the four  bins in projected separation bin $r_p$.  \label{fig:ChiSqrdFit} } 
		\end{center}
	\end{figure*}
	
	The overall goodness of fit $\chi^2$ values are listed in the Tables and plotted versus projected separation bin 
	in Figure~\ref{fig:ChiSqrdFit}; since the bins are width 0.1 and $\vtilde$ upper limits are respectively 5,6,7,7, there
	are respectively 50, 60, 70, 70 datapoints in the four separation bins.   
	The result seen in Figure~\ref{fig:ChiSqrdFit} is that all four separation bins show a common ordering, with 
	Newtonian gravity and $2e$ eccentricity distribution giving the best fit,, and MOND with flat $f(e)$ giving the worst. 
	Interestingly, all three Newtonian fits give a lower $\chi^2$ value than the best of the MOND models (also $f(e) = 2e$). 
	Therefore, since our three eccentricity models span the range expected for any reasonably smooth distribution, 
	the fact that we do not know the true eccentricity distribution seems not to be a serious problem for the future development
	of this test.  
	
	On inspection of the data and model histograms in Figs~\ref{fig:3BodyNewtonFlat} - \ref{fig:3BodyBanik2Ecc},
	it is clear that a substantial part of the difference is due to the region around velocity ratios $1.0 \le \tilde{v} \le 1.5$, 
	where all of the MOND fits overshoot the observed data values by a significant margin.  This region was 
	already highlighted as a key discriminant in \citetalias{Pittordis_2018}, since MOND predicts a substantially
	higher fraction of true binaries in this range compared to Newtonian gravity. Therefore, the predicted ``MOND excess" in this
	range appears {\bf not to be present in the data}, which will become a serious problem for MOND {\bf if} our modelling of
	the triple + flyby populations proves realistic. We are not yet confident in our modelling of the triple + flyby or other 
	populations, but hope to improve this in future work.    
	
	The formal significance of the preference for Newtonian gravity over MOND is very high, with the best
	Newtonian model ($2e$) {\newa having $\Delta\chi^2 \simeq -246, -164, -56, -60$  
		respectively} in the four bins compared to the best  MOND model (also $2e$);  the absolute $\chi^2$ values are acceptable for our best Newtonian model in the two wider bins,	but not in the two narrower bins, so the formal significance should not be overstated. 
	
	\subsection{Caveats and Limitations} 	
	Within the limited scope of the modelling above, the data shows a fairly strong preference for Newtonian gravity over our one MOND version. 
	However, there are several caveats and limitations which we discuss below. 
	\begin{enumerate} 
		\item	Perhaps the major caveat here is that our triple model is only a single realisation for each eccentricity distribution: for all other parameters we have taken a principle of ``maximum randomness" with uncorrelated masses, $e$ values, inclinations and orbit ratios between inner and outer orbits.  If the real triple population deviates from this simple assumption, then the predictions for the velocity distribution in triples will vary accordingly, and this will change the fit results.  However, we note that in order to improve the MOND fits, it is hard to change the shape of the model distribution for the pure binaries; 
		for pure binaries, assuming random phases and inclinations the model distribution is entirely determined given the $f(e)$ distribution function,  and we have already explored reasonable limiting cases for this.  Therefore, to improve the MOND fits we would need the triple+flyby models to produce fewer systems than our model in the range $1.0 \le \tilde{v} \le 1.5$, while maintaining the shape in the tail at $\tilde{v} > 1.5$; this may be challenging, but remains to be explored. 
		
		{\newa 
			\item Another limitation is that we have only explored a single realisation of MOND and a simplified
			treatment of the external field effect; there are many variants of 
			MOND-like models in the literature (see e.g. \citet{Barrientos_2018} for examples). While the 
			total acceleration law for a viable MOND is reasonably well constrained by the requirement to match 
			observed disk galaxy rotation curves, 
			the external field effect may vary between models; if the actual external field effect produces stronger convergence to
			Newtonian gravity than we have modelled above, then the observable MOND shift will be correspondingly weaker. } 
		
		
		\item 
		One other concern from a referee was that the apparent-magnitude limit of our sample leads to a distance-dependent
		mass limit in our selected systems; to leading order this should cancel since the estimated masses are included 
		in the definition of velocity ratio. To address this more directly, we constructed a volume-limited subsample 
		by discarding binaries where either star has absolute magnitude $M_G > 9.61$, corresponding to apparent $G > 17$ at 300 pc; 
		results for this volume-limited subsample are shown in Appendix~\ref{sec:vollim}, and are qualitatively similar to the full
		sample. 
		
		\item {\newa Another possible issue is that of distance-dependent selection. We  have explored cutting our above sample 
			in halves at the median distance of 210 pc, and fitting separately to the closer and more distant halves of the sample. 
			Results are shown in Appendix~\ref{sec:dcut}; the closer half shows distinctly larger $\chi^2$ differences, 
			while the distant half shows less clear distinction between models; but the
			sense of the differences is similar between closer and distant halves } .  
		
		\item {\newa As discussed above, a better understanding of the population of triples (and quadruples) is probably the
			key requirement to make the gravity test more secure. We could explore fitting with many variants of the triple model, 
			but this is computationally expensive due to the many degrees of freedom, and may lead to degeneracies. The best route forward appears to be a followup observing programme to directly detect the additional object(s) in triple systems. In a followup paper 
			\citep{Manchanda_2022} we have simulated various followup methods: the results are positive in that overall 
			some $80 - 95$ percent of third objects are detectable by a combination of GAIA astrometric accelerations, 
			deep seeing-limited imaging and followup speckle or coronagraphic imaging.  Essentially all main-sequence
			companions are detectable by one or more methods, while undetectable objects are primarily cool brown dwarfs
			at separations $\ga 25 \au$. }   
		
	\end{enumerate}

	\begin{table*}
		\caption{Number of candidate binaries in selected ranges of 
			projected separation and velocity ratio: data, and model fits for
			combined Binary (B), Triples (T) and FlyBys (F) populations,  
			for Newtonian gravity and Flat eccentricity distribution.  Rows are range of projected separation, as
			in Column 1.    Columns 2-4 are for all velocity ratios, 
			columns 5-7 for ratios $< 1.4$, and columns 8-10 for
			ratios between $1.1$ and $1.4$. Rightmost column 11 is the $\chi^2$ value of the fit.}
		\label{tab:3BodyNewtonFlat} 
		\centering 
		\small
		\begin{tabular}{lrrrrrrrrrrrrr} 
			\phantom{        } &  \multicolumn{4}{c}{$\vratio < 7$} & 
			\multicolumn{4}{c}{$\vratio < 1.4$} & 
			\phantom{        } &
			\multicolumn{3}{c}{$1.1 < \vratio < 1.4$} 
			\phantom{        } & \\
			\hline 
			$\rp$ range  &   Data & Fit(B) & Fit(T) & Fit(F)  & Data & Fit(B) & Fit(T) & Fit(F) & Data & Fit(B) & Fit(T) & Fit(F) & $\chi^2$ \\ 
			\hline 

			%
			$5-7.1 \kau$   & 2510 & 1279.7 & 1279.7 & 71.9  & 2054 & 1279.7 & 762.4 & 36.6 & 226 & 107.4 & 174.3 & 7.6  & 307.7 \\
			$7.1-10 \kau$  & 1960 & 917.3  & 917.3  & 206.8 & 1489 & 917.3  & 514.9 & 95.6 & 186 & 65.5  & 118.2 & 20.4 & 218.0 \\
			$10-14.1 \kau$ & 1499 & 704.9  & 704.9  & 115.8 & 1116 & 704.9  & 374.4 & 44.8 & 165 & 50.4  & 86.7  & 10.5 & 99.7  \\
			$14.1-20 \kau$ & 1307 & 571.2  & 571.2  & 173.9 & 902  & 571.2  & 297.1 & 54.6 & 137 & 46.1  & 70.3  & 14.2 & 99.7 \\
			\hline 
			\\ 
		\end{tabular}
		\caption{Same as Table~\ref{tab:3BodyNewtonFlat}, but for model orbits drawn from Tokovinin eccentricity distribution. } 
		\label{tab:3BodyNewtonTok} 
		
		\begin{tabular}{lrrrrrrrrrrrrr} 
			\phantom{        } &  \multicolumn{4}{c}{$\vratio < 7$} & 
			\multicolumn{4}{c}{$\vratio < 1.4$} & 
			\phantom{        } &
			\multicolumn{3}{c}{$1.1 < \vratio < 1.4$} 
			\phantom{        } & \\
			\hline 
			$\rp$ range  &   Data & Fit(B) & Fit(T) & Fit(F)  & Data & Fit(B) & Fit(T) & Fit(F) & Data & Fit(B) & Fit(T) & Fit(F) & $\chi^2$ \\
			\hline  
			$5-7.1 \kau$   & 2510 & 1281.7 & 1281.7 & 0.0   & 2054 & 1281.7 & 772.5 & 0.0  & 226 & 78.6 & 168.4 & 0.0  & 175.0 \\
			$7.1-10 \kau$  & 1960 & 919.2  & 919.2  & 162.1 & 1489 & 919.2  & 520.8 & 74.9 & 186 & 71.6 & 116.7 & 16.0 & 138.0 \\
			$10-14.1 \kau$ & 1499 & 692.2  & 692.2  & 129.4 & 1116 & 692.2  & 372.5 & 50.1 & 165 & 56.1 & 84.5  & 11.7 & 76.4  \\
			$14.1-20 \kau$ & 1307 & 559.9  & 559.9  & 181.8 & 902  & 559.9  & 298.4 & 57.0 & 137 & 43.2 & 68.4  & 14.8 & 68.2  \\
			\hline 
			\\ 
		\end{tabular}
		\caption{Same as Table~\ref{tab:3BodyNewtonFlat}, but model orbits drawn from $2e$ eccentricity distribution. } 
		\label{tab:3BodyNewton2Ecc} 
		\begin{tabular}{lrrrrrrrrrrrrr} 
			\phantom{        } &  \multicolumn{4}{c}{$\vratio < 7$} & 
			\multicolumn{4}{c}{$\vratio < 1.4$} & 
			\phantom{        } &
			\multicolumn{3}{c}{$1.1 < \vratio < 1.4$} 
			\phantom{        } & \\
			\hline 
			$\rp$ range  &   Data & Fit(B) & Fit(T) & Fit(F)  & Data & Fit(B) & Fit(T) & Fit(F) & Data & Fit(B) & Fit(T) & Fit(F) & $\chi^2$ \\
			\hline 
			%
			$5-7.1 \kau$   & 2510 & 1264.9 & 1264.9 & 0.0   & 2054 & 1264.9 & 776.3 & 0.0  & 226 & 99.9 & 165.4 & 0.0  & 109.8 \\
			$7.1-10 \kau$  & 1960 & 911.0  & 911.0  & 156.3 & 1489 & 911.0  & 530.9 & 72.2 & 186 & 65.4 & 115.2 & 15.4 & 94.6  \\
			$10-14.1 \kau$ & 1499 & 689.3  & 689.3  & 125.0 & 1116 & 689.3  & 380.1 & 48.4 & 165 & 56.3 & 84.5  & 11.3 & 58.0  \\
			$14.1-20 \kau$ & 1307 & 552.7  & 552.7  & 197.2 & 902  & 552.7  & 298.6 & 61.9 & 137 & 38.8 & 65.9  & 16.1 & 70.8  \\
			\hline 
		\end{tabular}
	\end{table*} 
	
	
	\begin{table*}
		\caption{As Table~\ref{tab:3BodyNewtonFlat}, but for realistic MOND gravity model
			and all orbits drawn from a flat eccentricity distribution.  Table shows number of candidate binaries in selected ranges of 
			projected separation and velocity ratio: data, and model fits for Binary(B), Triples (T) and FlyBys (F) populations.  
			Rows are range of projected separation, as
			in Column 1.    Columns 2-4 are for all velocity ratios, 
			columns 5-7 for ratios $< 1.4$, and columns 8-10 for
			ratios between $1.1$ and $1.4$. }
		\label{tab:3BodyBanikFlat}  
		\centering
		\small 
		\begin{tabular}{lrrrrrrrrrrrrr} 
			\phantom{        } &  \multicolumn{4}{c}{$\vratio < 7$} & 
			\multicolumn{4}{c}{$\vratio < 1.4$} & 
			\phantom{        } &
			\multicolumn{3}{c}{$1.1 < \vratio < 1.4$} 
			\phantom{        } & \\
			\hline 
			$\rp$ range  &   Data & Fit(B) & Fit(T) & Fit(F)  & Data & Fit(B) & Fit(T) & Fit(F) & Data & Fit(B) & Fit(T) & Fit(F) & $\chi^2$ \\
			\hline  
			
			%
			$5-7.1 \kau$   & 2510 & 1223.6 & 1223.6 & 296.1 & 2054 & 1201.6 & 702.0 & 150.7 & 226 & 241.3 & 179.8 & 31.4 & 531.3 \\
			$7.1-10 \kau$  & 1960 & 907.8  & 907.8  & 334.2 & 1489 & 885.6  & 486.5 & 154.4 & 186 & 196.8 & 131.9 & 32.9 & 432.8 \\
			$10-14.1 \kau$ & 1499 & 744.0  & 744.0  & 87.2  & 1116 & 733.0  & 375.7 & 33.8  & 165 & 154.3 & 105.0 & 7.9  & 202.0 \\
			$14.1-20 \kau$ & 1307 & 620.4  & 620.4  & 137.1 & 902  & 614.5  & 303.4 & 43.0  & 137 & 146.5 & 86.0  & 11.2 & 227.0 \\
			\hline 
			\\ 
		\end{tabular}
		\\ 
		\caption{Same as Table~\ref{tab:3BodyBanikFlat}, with realistic MOND gravity model, but model orbits
			drawn from Tokovinin eccentricity distribution. } 
		\label{tab:3BodyBanikTok} 
		\begin{tabular}{lrrrrrrrrrrrrr} 
			\phantom{        } &  \multicolumn{4}{c}{$\vratio < 7$} & 
			\multicolumn{4}{c}{$\vratio < 1.4$} & 
			\phantom{        } &
			\multicolumn{3}{c}{$1.1 < \vratio < 1.4$} 
			\phantom{        } & \\
			\hline 
			$\rp$ range  &   Data & Fit(B) & Fit(T) & Fit(F)  & Data & Fit(B) & Fit(T) & Fit(F) & Data & Fit(B) & Fit(T) & Fit(F) & $\chi^2$ \\
			\hline 
			
			%
			$5-7.1 \kau$   & 2510 & 1318.3 & 1318.3 & 70.8  & 2054 & 1299.2 & 772.9 & 36.1  & 226 & 313.2 & 190.6 & 7.5  & 462.0 \\
			$7.1-10 \kau$  & 1960 & 937.1  & 937.1  & 219.2 & 1489 & 914.2  & 510.3 & 101.3 & 186 & 188.1 & 132.1 & 21.6 & 323.8 \\
			$10-14.1 \kau$ & 1499 & 738.9  & 738.9  & 86.2  & 1116 & 719.7  & 378.4 & 33.4  & 165 & 135.7 & 100.1 & 7.8  & 178.7 \\
			$14.1-20 \kau$ & 1307 & 600.2  & 600.2  & 126.9 & 902  & 588.1  & 300.4 & 39.8  & 137 & 101.1 & 80.9  & 10.4 & 122.0 \\
			\hline 
			\\ 
		\end{tabular} 
		\\ 
		\caption{Same as Table~\ref{tab:3BodyBanikFlat} with MOND gravity model, but model orbits drawn from $f(e) = 2e$ eccentricity distribution.  } 
		\label{tab:3BodyBanik2Ecc} 
		\begin{tabular}{lrrrrrrrrrrrrr} 
			\phantom{        } &  \multicolumn{4}{c}{$\vratio < 7$} & 
			\multicolumn{4}{c}{$\vratio < 1.4$} & 
			\phantom{        } &
			\multicolumn{3}{c}{$1.1 < \vratio < 1.4$} 
			\phantom{        } & \\
			\hline 
			$\rp$ range  &   Data & Fit(B) & Fit(T) & Fit(F)  & Data & Fit(B) & Fit(T) & Fit(F) & Data & Fit(B) & Fit(T) & Fit(F) & $\chi^2$ \\
			\hline  
			
			%
			$5-7.1 \kau$   & 2510 & 1325.5 & 1325.5 & 0.0   & 2054 & 1286.8 & 791.0 & 0.0  & 226 & 248.0 & 193.7 & 0.0  & 355.8 \\
			$7.1-10 \kau$  & 1960 & 953.6  & 953.6  & 154.8 & 1489 & 922.2  & 530.8 & 71.5 & 186 & 180.9 & 132.2 & 15.2 & 258.2 \\
			$10-14.1 \kau$ & 1499 & 731.1  & 731.1  & 69.5  & 1116 & 713.1  & 381.7 & 26.9 & 165 & 129.1 & 98.2  & 6.3  & 114.1 \\
			$14.1-20 \kau$ & 1307 & 589.0  & 589.0  & 151.0 & 902  & 576.4  & 300.6 & 47.4 & 137 & 97.6  & 78.3  & 12.3 & 130.5 \\
			\hline 
		\end{tabular}
	\end{table*} 
	
	\subsection{Comparison with Hernandez et al (2022) study} 
	A recent paper by \citet{Hernandez_2022}, hereafter HCC22, 
	has also studied wide binaries 
	in GAIA DR3, and comes to a generally opposite conclusion that MOND is
	favoured over Newtonian gravity. This is mainly due to to a levelling-off
	of the observed RMS velocity difference; their binaries at $r_p > 0.01 \pc$ or
	$2 \kau$  show a near-flat 1D rms velocity difference at a value
	near $0.5 \kms$.  
	
	From the perspective of sample selection, 
	the main difference to this work is that HCC22 use much more 
	stringent constraints, e.g. $d < 130 \pc$ vs our $300\pc$, 
	RUWE $< 1.2$ instead of 1.4, 
	and tight constraints on the colour-magnitude diagram; these tight cuts
	are mainly motivated by removing contamination from triple systems, 
	clearly a desirable objective.   However, the downside is that
	their surviving sample for final analysis is very much smaller: HCC22 Figure 7 
	shows $< 200$ binaries with $2.06 < r_p < 20.6 \kau$, 
	compared to the 7276 used here between $5 < r_p < 20 \kau$.    
	Due to their small sample, detailed histogram-fits are not practical 
	and they use simple RMS velocity statistics; as is well known, RMS
	statistics are vulnerable to overestimation by 
	a small fraction of contaminants at values many times larger than
	the pure-population RMS. 
	
	In particular, the cuts used by HCC22 will likely not reject triples if the
	inner-orbit period is substantially longer than the 2.75 year DR3 timespan 
	(when the deviation from linear motion is too small to boost the RUWE value),
	and inner angular separation is below $\sim 0.5$ arcsec; 
	this leaves inner orbits from $\sim 5-50 \au$ 
	largely surviving their cuts, along with very faint third stars beyond 50 AU. 
	Such systems can cause a large boost on the observed $v_p$ of
	the wide system, as shown in simulations by \citet{Manchanda_2022}. 
	
	We note that there exist a
	handful of points in the upper-right corner of HCC22 Figure 7 which 
	have 1D relative velocity differences $> 5 \times$ higher than
	their Newtonian rms (blue line).  
	In our Newtonian orbit models, pure-binary systems have
	a maximum 2D velocity difference approximately $2.4 \times$ the population
	RMS at the same $r_p$; in 1D the maximum is unchanged but the 
	population RMS is $1/\sqrt{2}$ smaller,  so this limit 
	becomes $3.4 \times$; MOND with external field
	effect can boost this to near $4 \times$, but our simulations suggest 
	that systems above $5 \times$ the 1D rms are very unlikely to be  
	pure binaries, and suggestive of residual triples.  
	
\begin{figure*}[!htbp]
		\begin{center} 
			\includegraphics[width=0.45\linewidth]{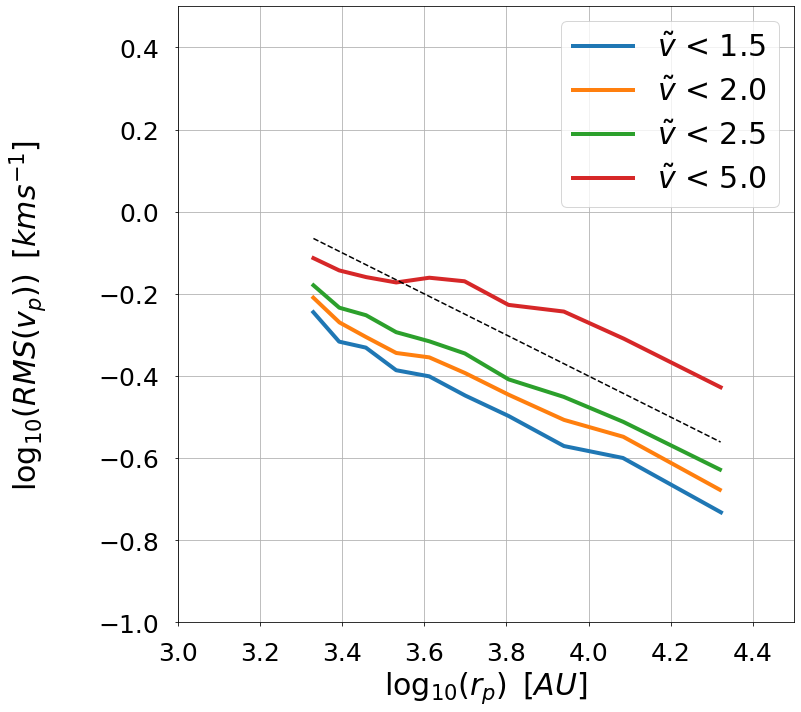} \, \,
			\includegraphics[width=0.45\linewidth]{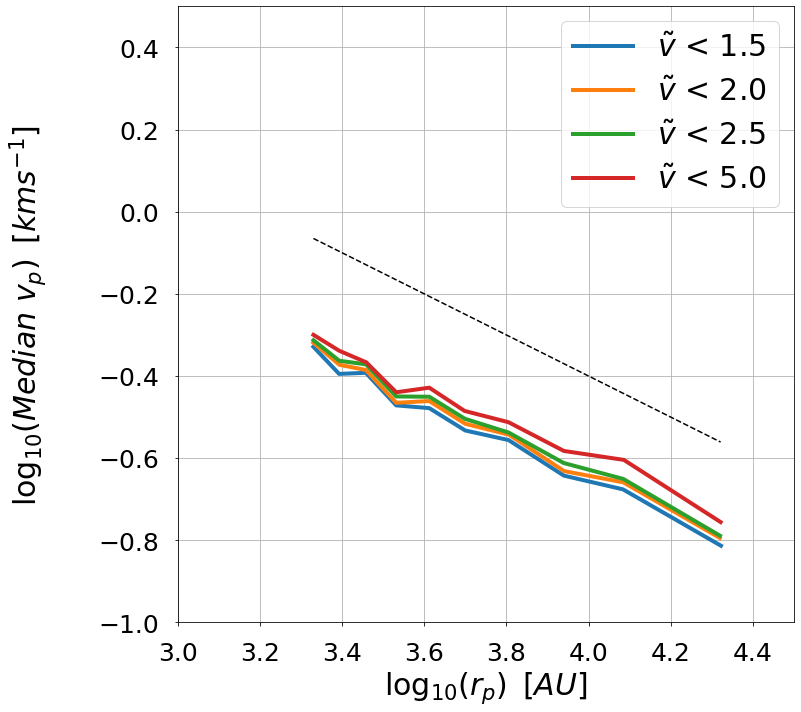} 
			\caption{{\bf Left panel:} The RMS sky-projected velocity difference vs $r_p$; the sample of candidate binaries with $2 < r_p < 20 \kau$ is divided into 10 $r_p$ bins of near-equal numbers; a threshold of $\vtilde < 5$, 2.5, 2 or 1.5 is applied (see legend), then the RMS in each bin is shown.  The dotted line indicates slope of $-0.5$, with arbitrary zero-point for visibility.   {\bf Right panel: } Same as left panel, but showing the median sky-projected $v_p$ instead of the RMS.}
			\label{fig:RMS_VP_logrp}  
		\end{center}
	\end{figure*}

	As a check, we have investigated RMS and median statistics
	of our sample over the range $2 < r_p < 20 \kau$ where the 
	MOND effect is claimed by HCC22.  Results are shown in Fig~\ref{fig:RMS_VP_logrp}.  Results are shown for several cuts on $\vtilde$ from 
	5 down to 1.5 to reject contaminants. 
	For the RMS values, there is a slow decline for $\vtilde < 5$ but a decline
	close to $r_p^{-0.5}$ for the smaller $\vtilde$ cuts which reject the tail.  
	The median statistic is more robust against contaminants, and all
	samples show a decline close to $r_p^{-0.5}$. 
	While our sample selection is different, 
	this suggests that population RMS velocity differences
	without a $\vtilde$ upper limit are strongly inflated by the
	tail of systems at $\vtilde > 2$ which cannot realistically be 
	pure binaries.  (If the MOND effect were so large as to allow pure
	binaries at $\vtilde > 2$, then the population mode (peak in
	our $\vtilde$ histograms) would show an obvious shift rightwards 
	with $r_p$,  but such a shift is not seen).   
	
	We conclude that triple-system contamination boosting the
	RMS velocity differences can plausibly account 
	for the MOND detection claimed by HCC22. 
	
	\newpage
	\section{Conclusions}
	\label{sec:conc} 
	Following on from previous modelling work in PS18, and our observational study from GAIA DR2 in PS19, we have selected a new sample of candidate wide binaries from the recent GAIA Early Data Release 3 (EDR3); our cleaned sample contains 73,159 candidate binaries with distance $\le 300 \pc$, projected separation $\le 50 \kau$ and projected velocity difference $\le 3 \kms$. 
	Comparisons with our previous PS19 sample and the independent binary sample of El-Badry et al (2021) show generally good consistency, after allowing for the different selection criteria used by El-Badry et al.   
	We then estimate masses for each component using a main-sequence mass/luminosity relation, and our key observable is the projected velocity difference in each binary, from the GAIA proper motions assuming a common distance, 
	divided by the velocity for a circular orbit at the observed projected separation; 
	the statistics of this ratio can be reliably predicted from simulations, assuming random inclinations and orbital phases, given
	an assumed eccentricity distribution.   
	
	We have then focused on the subset of 9063 candidate wide binaries with projected separations between $5 \kau$ to $20 \kau$, the most promising range for testing modified gravity; 
	{\newa applying a {\tt ruwe} cut  reduces the sample to 7276 candidates.}  
	In agreement with PS19, the distributions show a clear ``hump plus tail" shape, with a hump consistent with pure binaries and a tail to much larger velocity ratios. 
	Dividing the range into four separation bins, we then fit each observed distribution as an arbitrary mixture of binary plus triple populations; we repeat the fitting for Newtonian gravity and one specific MOND model, and three assumed $f(e)$ eccentricity distributions. 
	The fitting results show a clear preference for Newtonian gravity over MOND, with a high formal significance; this is 
	largely because the MOND model fits overshoot the observed number of systems in the range $1.0 \le \tilde{v} \le 1.5$. 
	Varying the model eccentricity distributions has little effect on this preference for standard gravity. However, this conclusion is only tentative at present due to using a somewhat over-simplified maximally-random model for the triple population. Further studies are clearly needed with a larger variety of models for the triple and flyby populations.  
	
	The prospects for future development of this test seem to be very good: GAIA EDR3 used only the first 33 months of GAIA survey data, while there are 7.5 years existing at present, and projected GAIA operations extend to 10 years, so the precision will substantially improve and allow even larger samples in future. This will be helpful,  but the statistical uncertainties seem to be already sub-dominant now; the major uncertainty at present appears to be in the modelling and fitting of the presumed triple and flyby population contributing the tail of high-velocity systems.  
	
	There are interesting prospects for observationally detecting or constraining the triples, including for example: searches for faint companions in deep sky surveys; adaptive-optics or speckle imaging to directly resolve inner binaries in triples; searches for astrometric accelerations or excess residuals with future GAIA data; and high-precision radial velocities from ground-based data, including possible time-dependence.  {\newa In a followup study \citep{Manchanda_2022} we have
		explored detection prospects for our simulated triples, with very promising conclusions: a high
		percentage of triples should be directly detectable as such. }  
	Such programs may require substantial amounts of telescope time to build up good statistics of triples and other contaminants,  but has a potentially large payoff in a decisive test of acceleration-based modified gravity models similar to MOND.  
	
	\section*{Acknowledgements} 
	We thank two anonymous referees for comments which significantly improved the paper. 
	We thank Indranil Banik for many helpful discussions, and  
	Zizhen Lin for contributions to the triple-system simulations. 
	CP has been supported by Colin Lee, Gary Barr, and Adel Szabo Malcolm (Legal \& General Investment Management), Peter Jackson (Outra) and Konstantin Orfenov (BNP Paribas). 
	
	\section*{DATA AVAILABILITY}
	The original data used in this paper is publicly available.
	The Gaia data can be retrieved through the Gaia archive
	(\url{https://gea.esac.esa.int/archive}). The binary catalogue CWB-EDR3
	will be made available on \url{https://zenodo.org} on acceptance of this paper. 
	
	
	
	%
	\clearpage  
	\bibliographystyle{aa_url} 
	\bibliography{Ref_PS22} 
	%
	
	\appendix
	\section{Comparison with PS19 sample from DR2}
	\label{sec:appa} 	
	
	Here we compare our new EDR3 sample with the wide binary sample from \citetalias{Pittordis_2019}, where very similar methods were used to find candidates. The CWB-DR2 sample of 24,282 candidates is compared via matching on RA \& Dec with $1 \arcsec$ search radius with the current WB-EDR3 sample of size 92,677 (before any cuts applied). In an ideal case all would have been recovered, since
	our distance and magnitude limits are expanded compared to \citetalias{Pittordis_2019}. 
	We found that of the 24,282 binaries in the cleaned sample of \citetalias{Pittordis_2019},  1696 or $\approx$ 7\% did not match a binary in WB-EDR3, before cleaning. This 7 percent of non-matches appears rather larger than expected, so we investigated the causes of these non-matched DR2 binary candidates. 
	
	On matching the individual stars, we found that 159 of the 1696 ``missing" binaries 
	had one or both stars not matching an EDR3 single star within a $1 \arcsec$ radius, presumably due to deblending or other
	reasons; since these are only 0.65 percent of the original CWB-DR2, we have not followed up this small minority.  
	For the other 1537 pairs where both DR2 stars matched an EDR3 star, but the DR2 binary did not appear in WB-EDR3, we re-ran a binary search with the only selection criterion as $r_p < 50 \kau$, resulting in 1518 binaries recovered, with binary parameters such as $v_p$, $r_p$ and parallax differences re-calculated from the EDR3 data. 
	
\begin{figure}[!htbp]
		\begin{center} 
			\includegraphics[width=\linewidth]{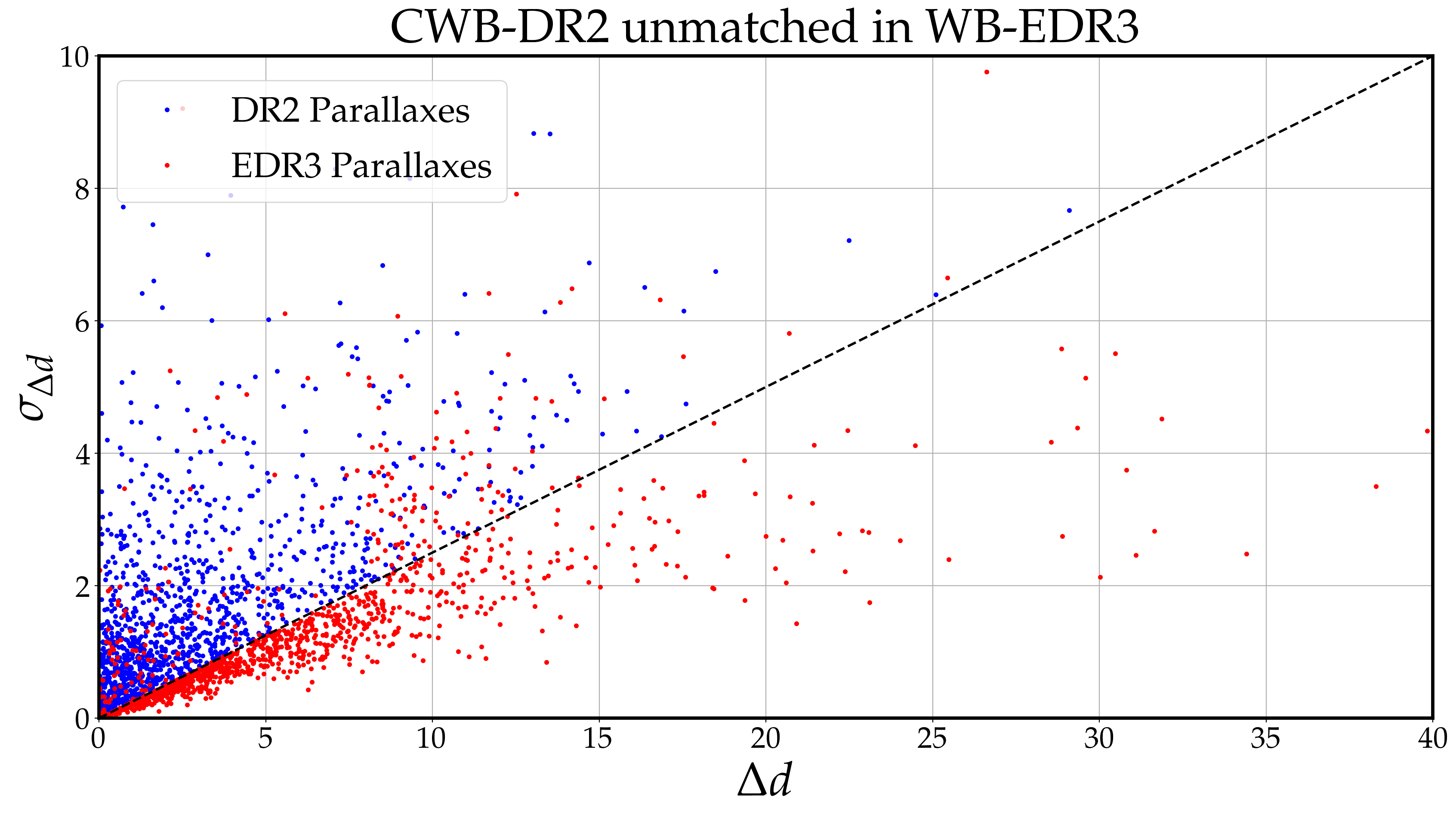} 
			\caption{Scatter plot for candidate binaries present in previous CWB-DR2 sample, but absent from WB-EDR3. The plot shows uncertainty in the difference in parallax distances $\sigma_{\Delta d}$ vs difference in distances $\Delta d$. The blue points show the values from CWB-DR2, while red points show updated values for the same systems from EDR3 data. 
				The sloping line is our selection criterion on this ratio. This plot shows that the most common reason for non-recovered DR2 binary candidates is those with larger $\Delta d$ in DR3. }
			\label{fig:Edr3Gdr2_DeltaDistanceDiff}  
		\end{center}
	\end{figure}

	We then investigated which of the cut(s) were failed for each of these 1518 pairs. 
	After applying star-quality cuts in EDR3 (using the same method as described in section \ref{sec: Data quality cuts}), 307 failed for one or both stars, while 1211 binaries survived. 
	We then found that the most common reason for failure was the criterian $ \vert d_1 - d_2 \vert \leq 4\sigma_{\Delta d}$; 
	this resulted in 1153 failures (1,008 after applying star-quality cuts), as shown in Figure \ref{fig:Edr3Gdr2_DeltaDistanceDiff} .  
	The second most common reason was $v_p < 3 \kms$ , which
	accounted for another 190 (183 after applying star-quality cuts) binaries exceeding the cut given the EDR3 values. Table \ref{tab:EDR3compCWBGDR2NotInWBEDR3} shows the number of Passed \& Failed for each of the criteria. 
	
	From that Table, it is clear that the main reason for non-recovery of PS19 binaries is pairs with parallax differences below our threshold in DR2, but above threshold in EDR3.  Our  $4\sigma$ threshold is fairly generous, so this is rather unexpected for Gaussian errors. This 
	might perhaps be explained by a population of hierarchical triples for which orbital motion results in substantially
	non-Gaussian parallax errors in the standard 5-parameter fits. Future data from DR4 or beyond may help to 
	clarify this; for the present, we do not use these ``missing" binaries in the rest of this paper. 
	
\begin{figure}[!htbp]
		\begin{center} 
			\includegraphics[width=\linewidth]{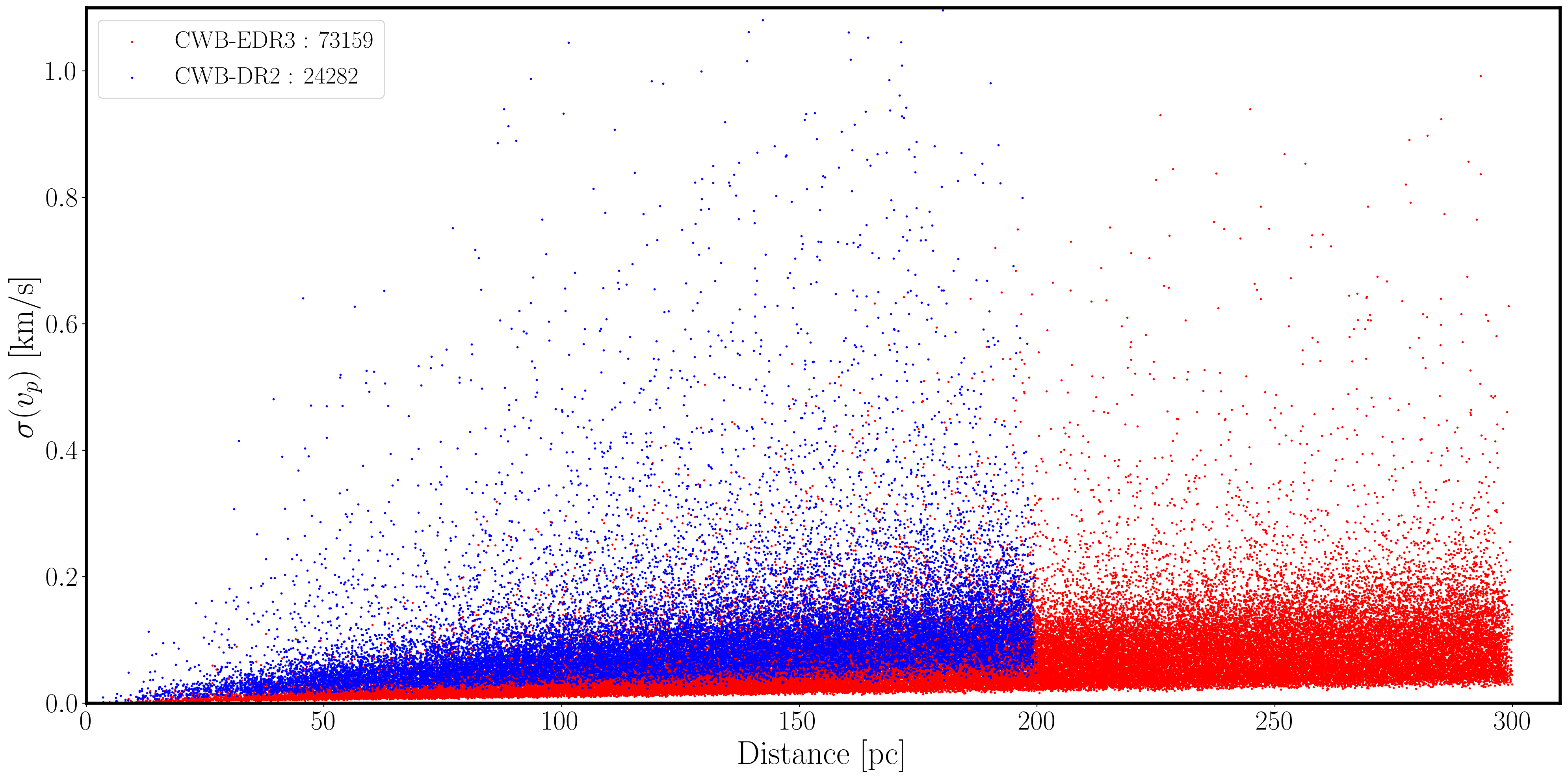} 
			\caption{Scatter plot of rms velocity uncertainty $\sigma(\vp)$ versus
				mean distance for CWB-EDR3 (red) and CWB-DR2 (blue) binaries surviving all cuts.}
			\label{fig:dsigv_wGDR2} 
		\end{center}
	\end{figure}

\begin{figure*}
		\begin{center} 
			\includegraphics[width=\linewidth]{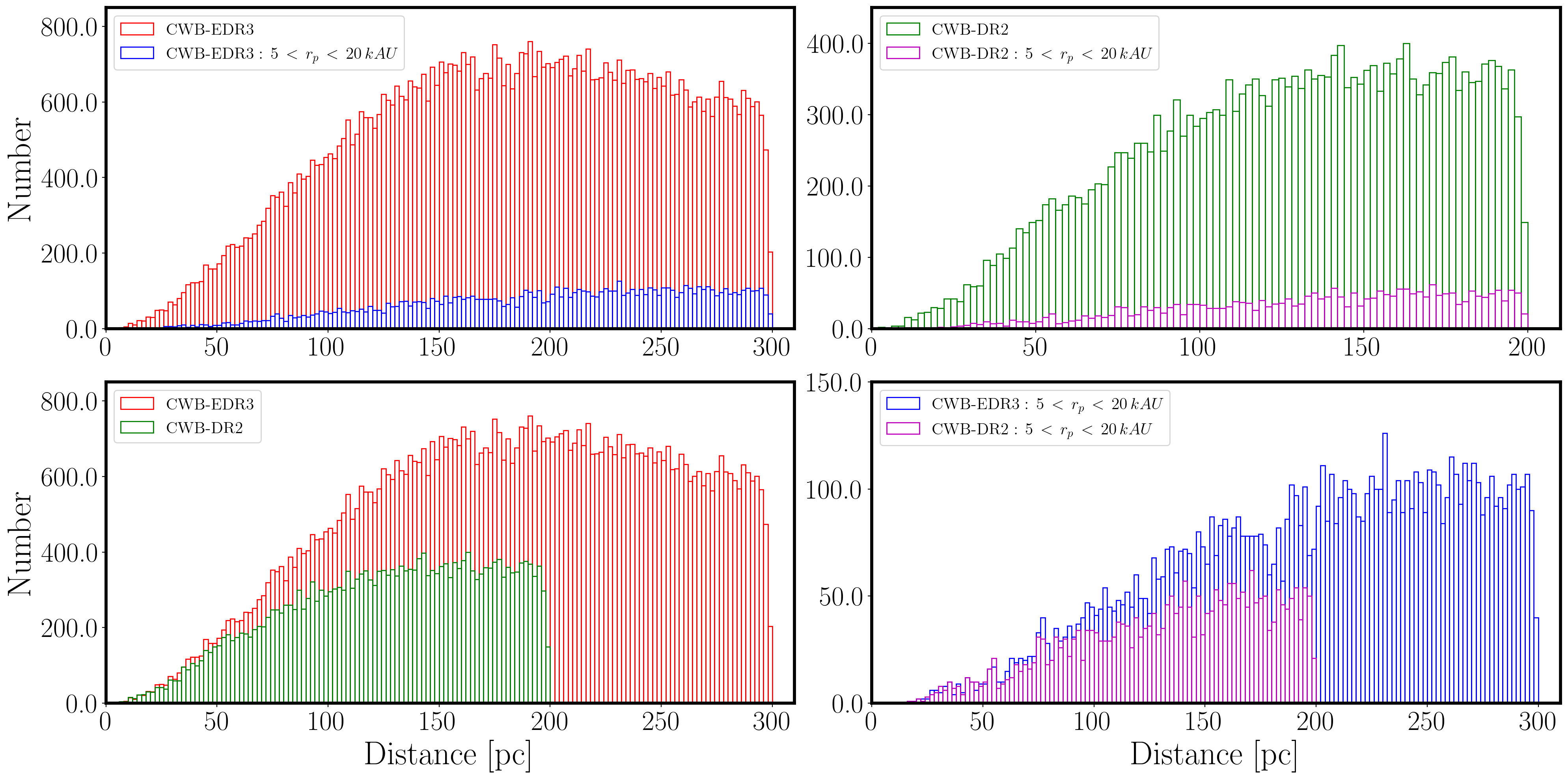} 
			\caption{Histograms of average distance for the candidate binaries. 
				(Top Right) histogram shows all 73,159 CWB-EDR3, also showing the subset with $5 < \rp < 20 \kau$ comparing between EDR3 and GDR2. (Top Left) same as the (Top Right) histogram but for CWB-DR2. (Bottom Right) compares between CWB-EDR3 and CWB-DR2, and same for the (Bottom Right) but for the subset of $5 < \rp < 20 \kau$ .   }
			\label{fig:dhist} 
			\includegraphics[width=\linewidth]{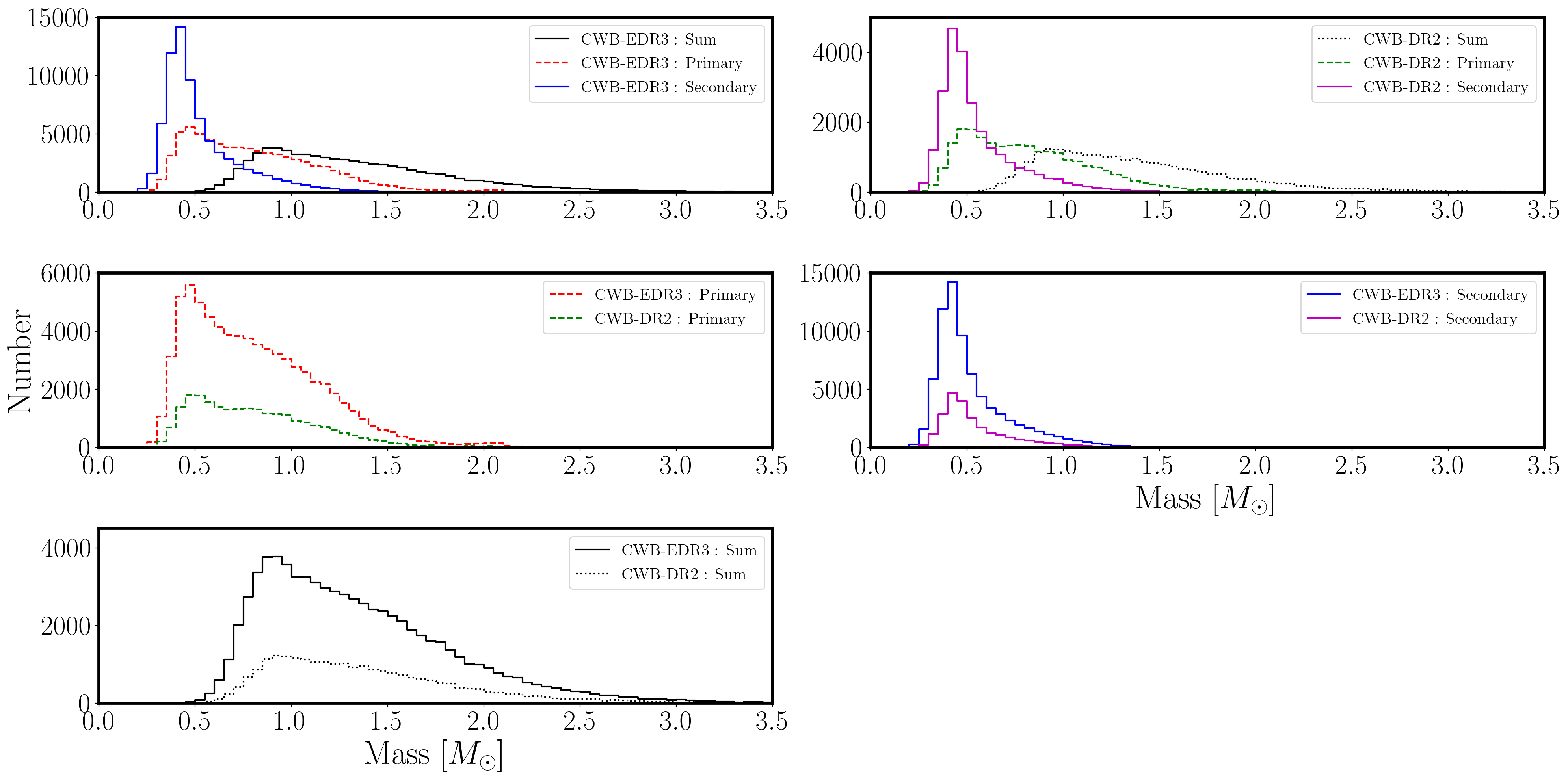} 
			\caption{Histograms of estimated masses for the candidate binaries. 
				For the CWB-EDR3 sample, the black solid line shows combined system mass; dashed red line
				shows the primary (more massive) star, and blue line shows
				the secondary star. For the CWB-DR2 sample,the dashed black line shows combined system mass; dashed green line shows the primary (more massive) star, and magenta line shows the secondary star. The rest of the histograms compared the combined system mass, primary and secondary mass between CWB-EDR3 and CWB-DR2. } 
			\label{fig:mhist} 
		\end{center}
	\end{figure*}

	Additionally, in Figs.~\ref{fig:dsigv_wGDR2} to \ref{fig:mhist} we show comparisons of velocity errors, distances and masses 
	between PS19 and the new sample here, showing results as expected. 
	
	\vfill\eject

	\begin{table}[!htbp]
		\caption{Reasons for subsample of PS19 binaries not recovered in 
			current CWB-EDR3 sample: number of Passed/Failed for each criterion of the  candidate binaries. Values in brackets are without  star-quality cuts.}
		\label{tab:EDR3compCWBGDR2NotInWBEDR3}  
		\centering 
		\small
		\begin{tabular}{l l c}
			\hline
			Criterion  &   Passed  & Failed    \\ 
			\hline
			$r_p < 50 \kau$  & 1,211 (1,518) &  0   \\
			$\vert d_1 - d_2 \vert < 8 \pc$ & 880 (978) &  331 (540)   \\
			$ \vert d_1 - d_2 \vert \leq 4\sigma_{d}$ & 203 (365) & 1,008 (1,153)     \\
			$\Delta v_p < 3 \kms $ & 1,073 (1,328) & 138 (190)     \\
			Combination of all criteria above & 0 & 1,211 (1,518)  \\
			\hline
		\end{tabular}
	\end{table} 
	
	\clearpage
	
	\clearpage
	\section{Volume limited subample}
	\label{sec:vollim} 
\begin{figure*} 
		\begin{center} 
			\includegraphics[width=\linewidth]{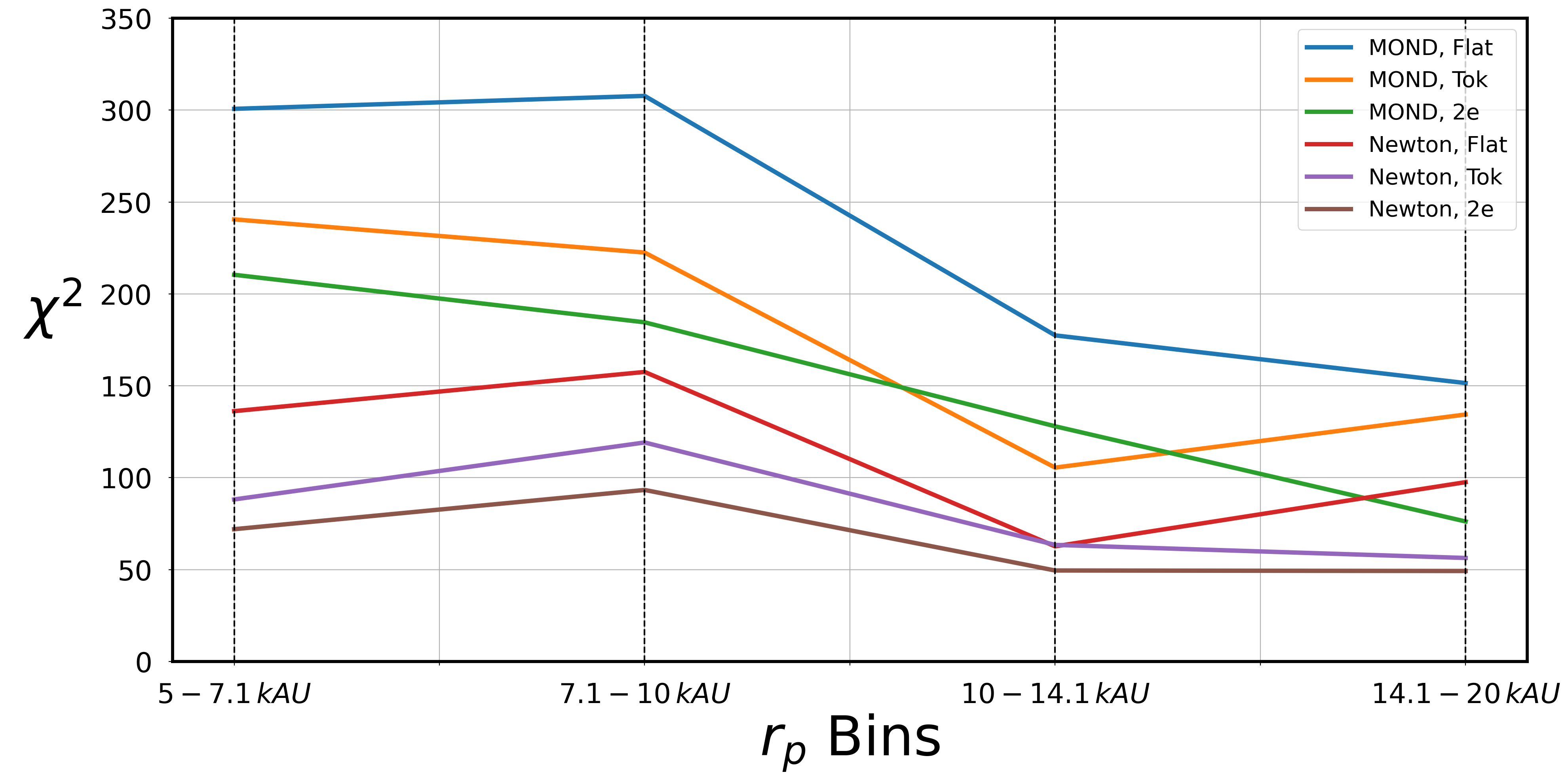} 
			\caption{ Same as Figure~\ref{fig:ChiSqrdFit}, except with the Volume-limit cut applied, showing a plot comparing the $\chi^2$ fitting results for each of the six orbit models (lines labelled in legend) against the four  bins in projected separation bin $r_p$. } \label{fig:ChiSqrdFit_VolLimit}
		\end{center}
	\end{figure*}

\begin{figure*} 
		\begin{center} 
			\includegraphics[width=\linewidth]{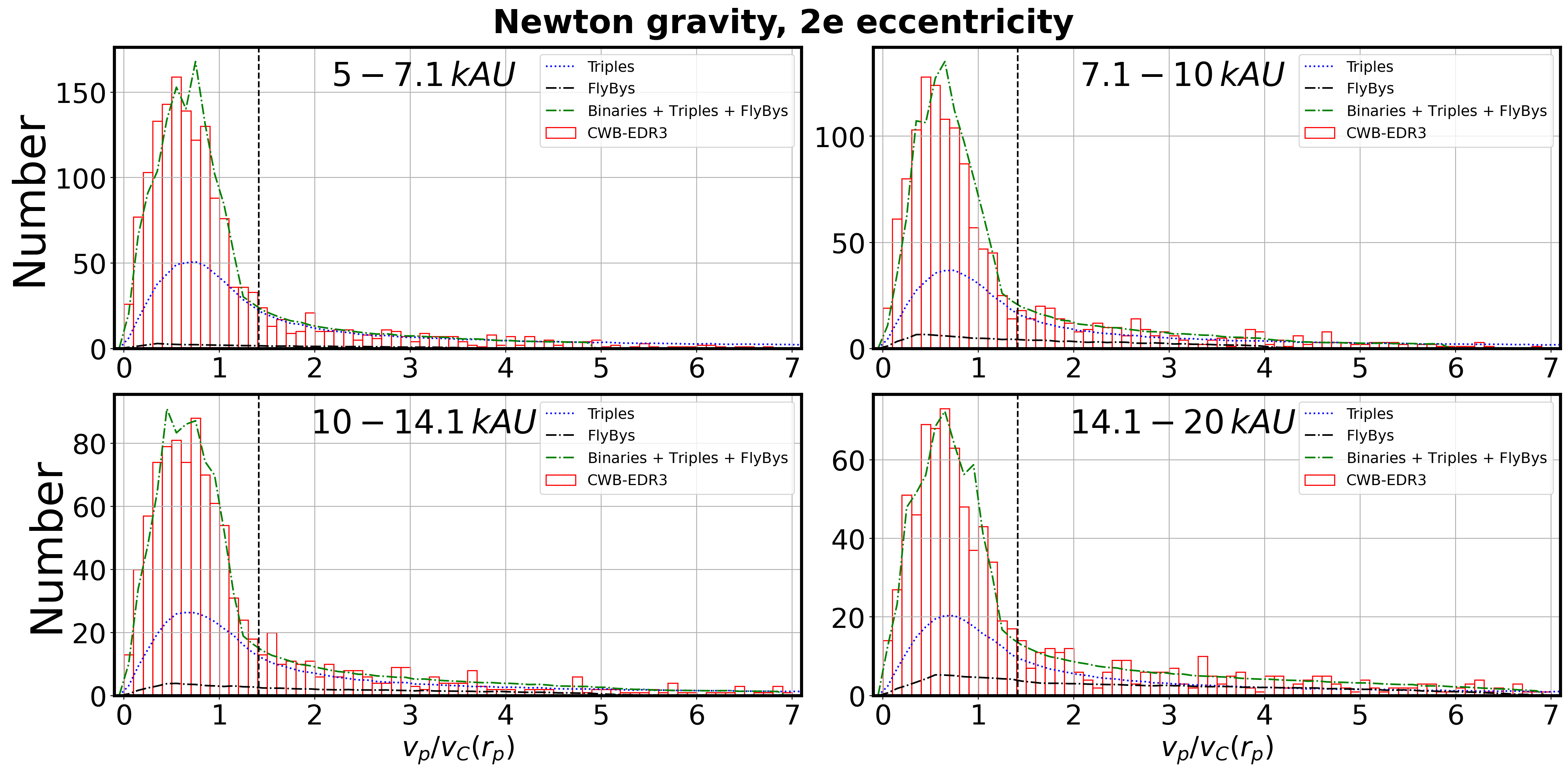} 
			\caption{Same as Figure~\ref{fig:3BodyNewton2Ecc} with Newtonian gravity and $f(e) = 2e$ eccentricity distribution, except with the Volume-limit cut applied.  \label{fig:3BodyNewton2Ecc_VolLimit}  }  
			
			\includegraphics[width=\linewidth]{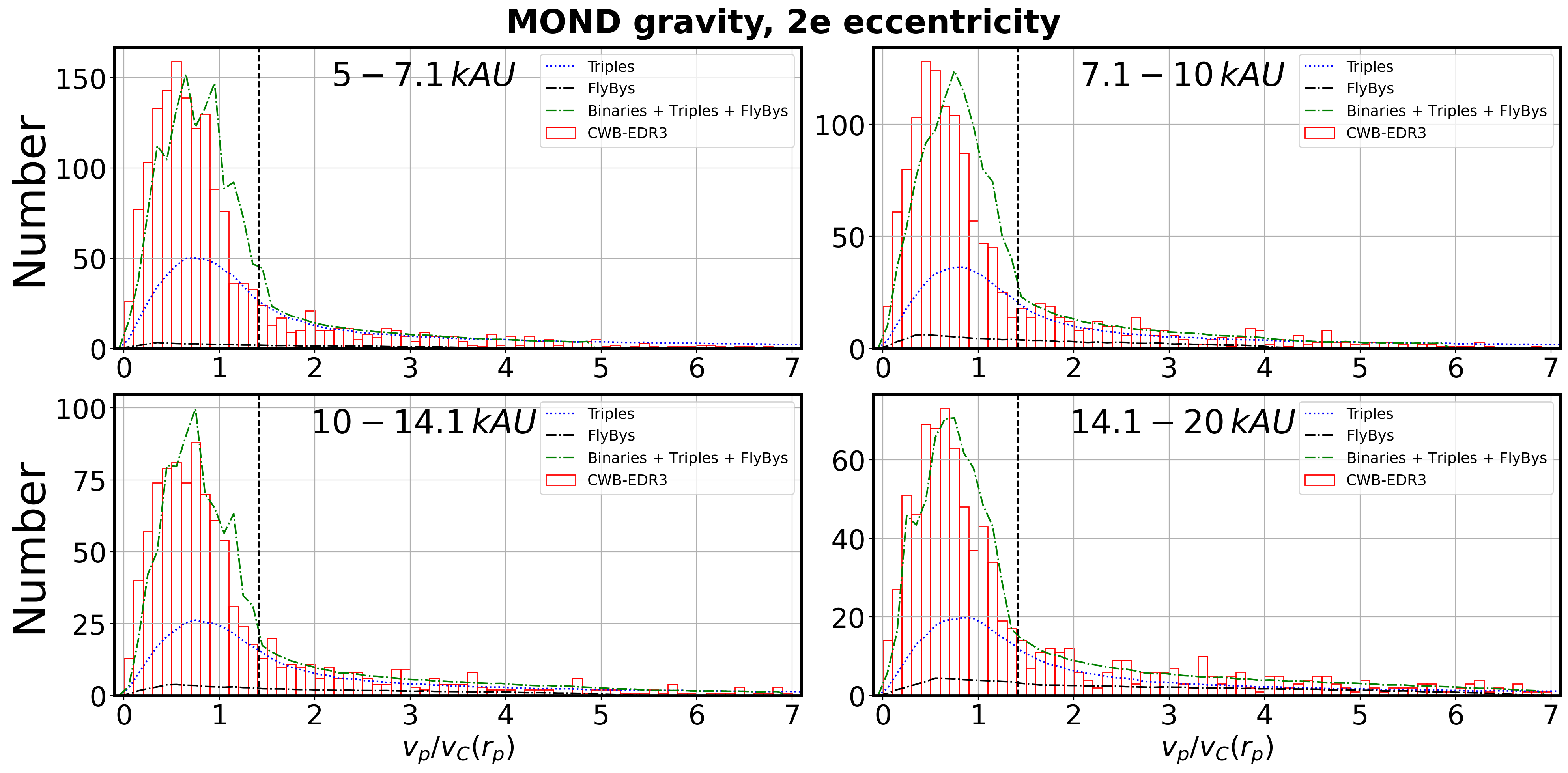} 
			\caption{ Same as Figure~\ref{fig:3BodyBanik2Ecc} with realistic MOND gravity model and and $f(e) = 2e$ eccentricity distribution, except with the Volume-limit cut applied.} 
			\label{fig:3BodyBanik2Ecc_VolLimit}
		\end{center}
	\end{figure*}

	A concern was raised by a referee that the $G < 17$ apparent magnitude limit leads to a distance-dependent selection, i.e.
	loss of low-mass systems at larger distance; to leading order
	we would not expect any effect since the system masses are scaled out in the velocity ratio $\vtilde$.  To test for any effect, 
	we applied a volume-limit to our CWB-EDR3 sample, by discarding all binaries where either star would be fainter than $G > 17$  if at 300$\pc$, which corresponds to a Gaia absolute magnitude cut $M_G < 9.61$. This reduces the sample from 77,159 to 44,285 candidate binaries, hereafter labelled as CWB-EDR3-VolLim. We also repeated the fit procedure as described in Section~\ref{sec:comp}, fitting the two gravity and three $f(e)$ models to the volume-limited subsample.  
	The resulting $\chi^2$ values are shown in  \ref{fig:ChiSqrdFit_VolLimit}, and follow essentially the
	same rank-ordering as the full sample.  
	For brevity we only show histograms (see Figures ~\ref{fig:3BodyNewton2Ecc_VolLimit} and ~\ref{fig:3BodyBanik2Ecc_VolLimit}), and tables (Tables ~\ref{tab:3BodyNewton2Ecc_VolLimit} and ~\ref{tab:3BodyBanik2Ecc_VolLimit}) for both gravity models with the $f(e) = 2e$ eccentricity distribution, i.e. the best eccentricity model 
	for both Newton and MOND. 
	
	For the volume-limited subsample, we note that the fit $\chi^2$ values are marginally improved 
	over the full subsample, as shown in Tables ~\ref{tab:3BodyNewton2Ecc_VolLimit} and ~\ref{tab:3BodyBanik2Ecc_VolLimit}; though the $\chi^2$ differences between Newtonian and MOND are reduced, as expected given the $0.67 \times$ reduction in
	sample size hence relatively larger Poisson errors. 
	\clearpage
	
	\begin{table*}
		\caption{Number of candidate binaries in selected ranges of 
			projected separation and velocity ratio: data, and model fits for
			combined Binary (B) and Triples (T) populations,  
			for Newtonian gravity and Flat eccentricity distribution.  Rows are range of projected separation, as
			in Column 1.    Columns 2-4 are for all velocity ratios, 
			columns 5-7 for ratios $< 1.4$, and columns 8-10 for
			ratios between $1.1$ and $1.4$. Rightmost column 11 is the $\chi^2$ value of the fit.
			Same as Table~\ref{tab:3BodyNewtonFlat}, but model orbits drawn from $2e$ eccentricity distribution. }
		\label{tab:3BodyNewton2Ecc_VolLimit} 
		\small
		\begin{tabular}{lrrrrrrrrrrrrr} 
			\phantom{        } &  \multicolumn{4}{c}{$\vratio < 7$} & 
			\multicolumn{4}{c}{$\vratio < 1.4$} & 
			\phantom{        } &
			\multicolumn{3}{c}{$1.1 < \vratio < 1.4$} 
			\phantom{        } & \\
			\hline 
			$\rp$ range  &   Data & Fit(B) & Fit(T) & Fit(F)  & Data & Fit(B) & Fit(T) & Fit(F) & Data & Fit(B) & Fit(T) & Fit(F) & $\chi^2$ \\
			\hline 
			$5-7.1 \kau$   & 1594 & 777.10 & 777.10 & 51.93  & 1268 & 777.10 & 476.89 & 26.43 & 148 & 63.07 & 101.61 & 5.51  & 72.00 \\
			$7.1-10 \kau$  & 1304 & 597.01 & 597.01 & 138.72 & 988  & 597.01 & 347.90 & 64.09 & 117 & 45.39 & 75.48  & 13.66 & 93.35 \\
			$10-14.1 \kau$ & 1007 & 460.08 & 460.08 & 97.35  & 746  & 460.08 & 253.69 & 37.70 & 109 & 38.23 & 56.40  & 8.82  & 49.52 \\
			$14.1-20 \kau$ & 874  & 355.58 & 355.58 & 163.28 & 592  & 355.58 & 192.13 & 51.23 & 96  & 32.23 & 42.43  & 13.34 & 49.25 \\
			\hline 
		\end{tabular}
		\\
		\\
		\caption{As Table~\ref{tab:3BodyNewtonFlat}, but for realistic MOND gravity model
			and all orbits drawn from a flat eccentricity distribution.  Table shows number of candidate binaries in selected ranges of 
			projected separation and velocity ratio: data, and model fits for binary(B) and triples (T) populations.  
			Rows are range of projected separation, as
			in Column 1.    Columns 2-4 are for all velocity ratios, 
			columns 5-7 for ratios $< 1.4$, and columns 8-10 for
			ratios between $1.1$ and $1.4$.
			Same as Table~\ref{tab:3BodyBanikFlat} with MOND gravity model, but model orbits drawn from $f(e) = 2e$ eccentricity distribution.  } 
		\label{tab:3BodyBanik2Ecc_VolLimit} 
		\begin{tabular}{lrrrrrrrrrrrrr} 
			\phantom{        } &  \multicolumn{4}{c}{$\vratio < 7$} & 
			\multicolumn{4}{c}{$\vratio < 1.4$} & 
			\phantom{        } &
			\multicolumn{3}{c}{$1.1 < \vratio < 1.4$} 
			\phantom{        } & \\
			\hline 
			$\rp$ range  &   Data & Fit(B) & Fit(T) & Fit(F)  & Data & Fit(B) & Fit(T) & Fit(F) & Data & Fit(B) & Fit(T) & Fit(F) & $\chi^2$ \\
			\hline  
			$5-7.1 \kau$   & 1594 & 808.26 & 808.26 & 59.43  & 1268 & 774.30 & 482.34 & 30.24 & 148 & 129.05 & 118.11 & 6.30  & 210.44 \\
			$7.1-10 \kau$  & 1304 & 625.55 & 625.55 & 129.04 & 988  & 612.44 & 348.16 & 59.62 & 117 & 104.88 & 86.69  & 12.71 & 184.57 \\
			$10-14.1 \kau$ & 1007 & 478.47 & 478.47 & 96.63  & 746  & 467.04 & 249.80 & 37.42 & 109 & 81.41  & 64.24  & 8.76  & 128.00 \\
			$14.1-20 \kau$ & 874  & 373.91 & 373.91 & 137.54 & 592  & 373.91 & 190.82 & 43.15 & 96  & 59.46  & 49.69  & 11.23 & 76.19 \\
			\hline 
		\end{tabular}
	\end{table*}
	
	\section{Distance-cut subsamples}
	\label{sec:dcut} 
\begin{figure*} 
		\begin{center} 
			\includegraphics[width=\linewidth]{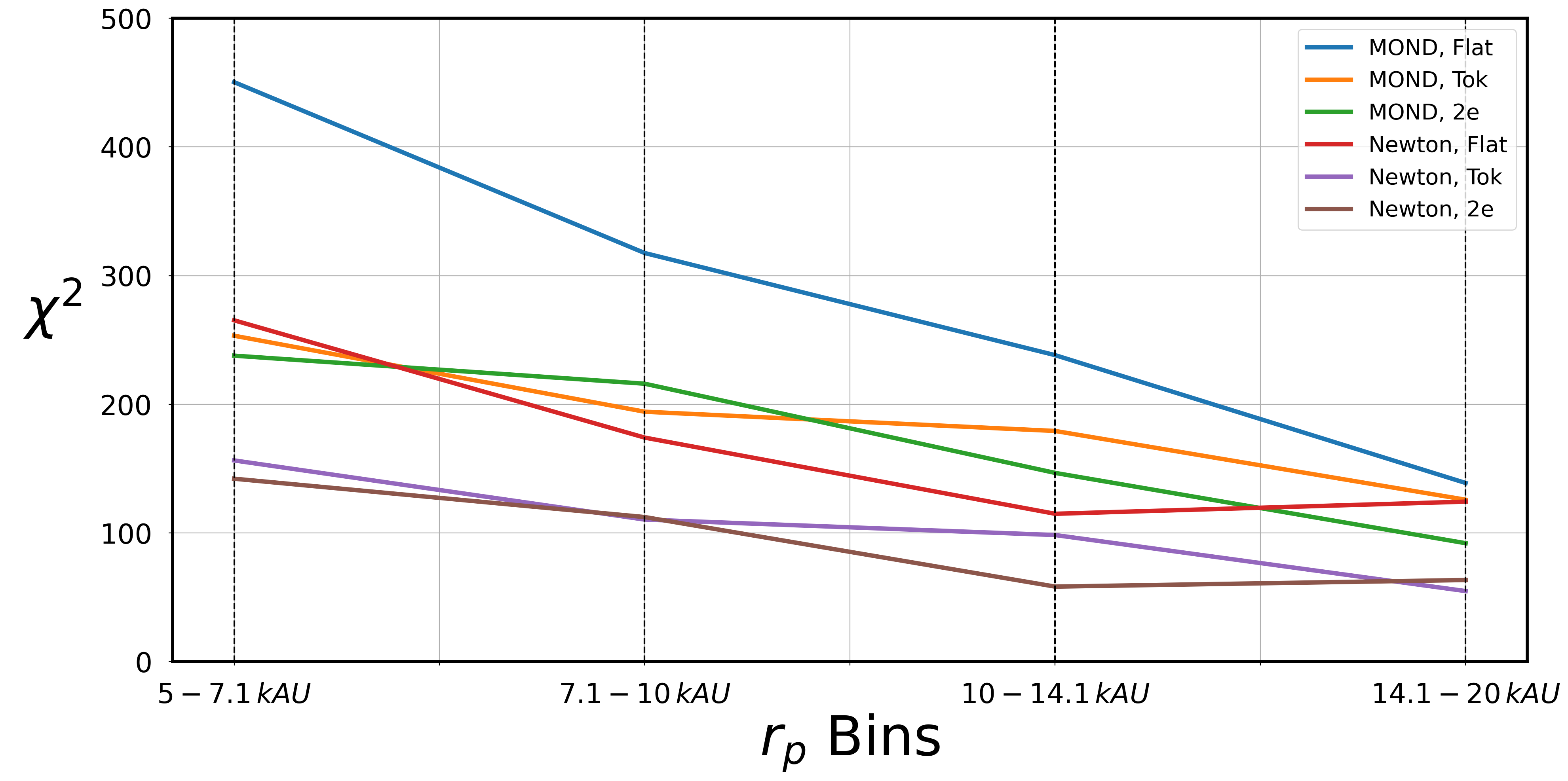} 
			\caption{ Same as Figure~\ref{fig:ChiSqrdFit}, for the near half of the sample at $d < 210 \pc$; the plot compares the $\chi^2$ fitting results for each of the six orbit models (lines labelled in legend) against the four bins in projected separation  $r_p$. } \label{fig:ChiSq_nearhalf}
		\end{center}
	\end{figure*}

\begin{figure*} 
		\begin{center} 
			\includegraphics[width=\linewidth]{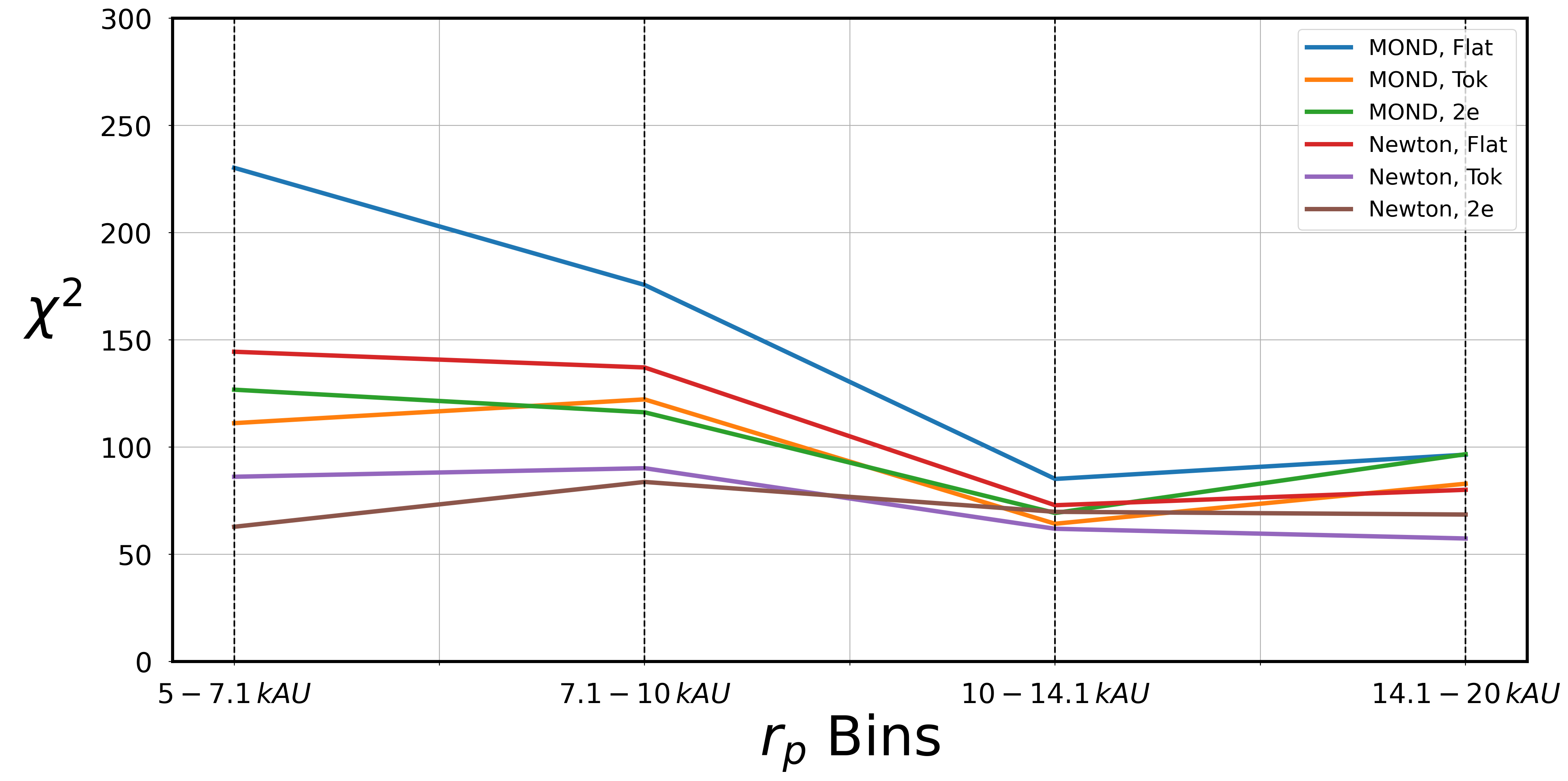} 
			\caption{ Same as Figure~\ref{fig:ChiSqrdFit}, for the far half of the sample at $210 \pc < d < 300 \pc$; the plot compares the $\chi^2$ fitting results for each of the six orbit models (lines labelled in legend) against the four bins in projected separation  $r_p$. } \label{fig:ChiSq_farhalf}
		\end{center}
	\end{figure*}

	Here we investigate whether there is a systematic distance-dependent effect; we have cut the wide-binary sample 
	in half at the median distance of $210 \pc$, then repeated the fit procedure from Section~\ref{sec:comp} 
	separately for the near half ($d < 210 \pc$) and the far half ($210 < d < 300 \pc$) of the sample. 
	The results for the $\chi^2$ values are shown in Figure~\ref{fig:ChiSq_nearhalf} for the near half, and Figure~\ref{fig:ChiSq_farhalf}  for the distant half. 	 
	
	\clearpage
	
	\label{lastpage}   

	----------------------------------------------------------------

\end{document}